\documentclass[3p,12pt]{elsarticle}
\usepackage{lineno}
\modulolinenumbers[5]
\usepackage{graphicx}
\usepackage{amssymb}
\usepackage{bm,amsmath}
\usepackage{caption}
\usepackage{subeqnarray}
\usepackage{multirow}
\usepackage{lscape}
\usepackage{rotating}
\usepackage{subcaption}
\usepackage{lineno}

\usepackage{color,ulem}

\def\x{{\bm x}}
\def\bxi{{\bm \xi}}

\def\bxi{{\bm{\xi}}}

\def\be{\begin{equation}}
\def\ee{\end{equation}}

\journal{Journal of computational physics}

\begin{document}
\title{A comparative study of discrete velocity methods for rarefied gas flows}
\author{Peng Wang\fnref{label1}}%
\author{Minh-Tuan Ho\fnref{label1}}%
\author{Lei Wu\fnref{label1}}%
\author{Zhaoli Guo \fnref{label2}}
\author{Yonghao Zhang\corref{cor1}\fnref{label1}}%
 \ead{yonghao.zhang@strath.ac.uk}
\address[label1]{James Weir Fluids Laboratory, Department of Mechanical and Aerospace Engineering, University of Strathclyde, Glasgow G1 1XJ, UK}
\address[label2]{State Key Laboratory of Coal Combustion, Huazhong University of Science and Technology, Wuhan 430074, China}
\cortext[cor1]{Corresponding author}
\date{\today}

\begin{abstract}
In the study of rarefied gas dynamics, the discrete velocity method (DVM) has been widely employed to solve the gas kinetic equations.
Although various versions of DVM have been developed, their performance, in terms of accuracy and computational efficiency,
is yet to be compreheively studied in the whole flow regime. Here, the traditional third-order time-implicit Godunov DVM (GDVM) and the recently developed discrete unified gas-kinetic scheme (DUGKS) are analysed in finding steady-state solutions of the force-driven
Poiseuille and lid-driven cavity flows. With the molecular collision and free streaming being treated simultaneously, the DUGKS preserves the second-order
accuracy in the spatial and temporal discretizations in all flow regimes. Towards the hydrodynamic flow regime, the DUGKS is not only faster
than the GDVM when using the same spatial mesh, but also requires less spatial resolution than that of the GDVM to achieve the same numerical accuracy.
From the slip to free molecular flow regimes, however, the DUGKS is slower than the GDVM, due to the complicated flux evaluation and the time step is less
than the maximum effective time step of the GDVM. Therefore, the DUGKS is preferable for problems involving different flow regimes, particularly when the hydrodynamic flow regime is dominant. For highly rarefied gas flows, if the steady-state solution is concerned, the implicit DVM, which can boost the convergence  significantly, is a better choice.
\end{abstract}


\maketitle
\linenumbers

\section{\label{sec:level1}Introduction}


Multi-scale flows, where different temporal and spatial scales are associated with different physical processes, are involved in many engineering problems. The modeling and simulation of these flows remains a research challenge. The gas flow at different scales can be categorized by the Knudsen number ($Kn$), defined as the ratio of the mean free path of gas molecules to the characteristic flow length. It is well recognized that
the computational fluid dynamics based on the Navier-Stokes (NS) equations and the direct simulation Monte Carlo (DSMC) method~\cite{bird1994molecular}
are two dominant (efficient and accurate) methods for the simulation of the hydrodynamic ($Kn< 10^{-3}$) and rarefied gas  (transition, $0.1<Kn< 10$; free molecular,
$Kn> 10$) flows, respectively. However, in the slip regime ($10^{-3}<Kn< 0.1$), the NS solvers and the DSMC method become either inaccurate or inefficient:
the NS equations are inappropriate to describe rarefied (non-equilibrium) gas flows because they are derived based upon the near equilibrium hypothesis, while the particle nature of DSMC method restricts its application in near hydrodynamic regime~\cite{alexander1998cell}, as the temporal and spatial resolutions must be smaller than the molecular collision time and mean free path, respectively.
Therefore, for multi-scale gas flows, it is intuitive to use continuum-particle hybrid methods that solve the flow fields in different regimes by appropriate solvers~\cite{o1995molecular,weinan2003heterogeneous,werder2005hybrid,borg2013fluid,borg2015hybrid}.
However, hybrid methods may encounter great difficulties for flows with a continuous and complex variation
of flow physics~\cite{radtke2013efficient}.

The Boltzmann equation is a fundamental model for dilute gas flows in all flow regimes, which uses a one-particle velocity distribution function defined
in a six-dimensional phase space to describe the system state. Near the hydrodynamic regime, the NS equations can be derived through the Chapman-Enskog
expansion. Thus, the modeling of multi-scale gas flows is not a problem, but the numerical solution of the Boltzmann equation remains a grand research challenge. In the past two decades, deterministic numerical methods
have been developed to solve the Boltzmann equation~\cite{mieussens2014survey}, most of which are based on the discrete velocity
method (DVM)~\cite{broadwell1964study,broadwell1964shock,yang1995rarefied,mieussens2000discrete,li2009gas,meng2011accuracy,meng2013lattice} that approximates
the continuous molecular velocity space by discrete velocity points, so that the resulting equations can be solved numerically~\cite{yang1995rarefied}. Many full Boltzmann solvers~\cite{sone1989temperature,ohwada1993structure,aristov2001direct,tcheremissine2005direct, mouhot2006fast,wu2013deterministic, wu2014solving, wu2016non}, especially the fast spectral method~\cite{wu2014solving,wu2015kinetic,wu2015fast, wu2016non, wu2016sound},
provide accurate numerical results, which can serve as reference solutions. However, the high computational cost in calculating the complicated  collision operator makes them impractical for many applications~\cite{mieussens2014survey}.
Therefore, the Boltzmann equation is usually replaced by simplified kinetic model equations, such as the BGK~\cite{bhatnagar1954model}, ES~\cite{holway1966new},
and Shakhov~\cite{shakhov1968generalization} models. And most of DVMs are developed for these Boltzmann model equations.

In the traditional DVM, the Boltzmann model equation is explicitly solved through the operator splitting method~\cite{aristov2001direct},
where the time step and cell size are limited by the mean collision time and  mean free path of gas molecules, respectively.
Consequently, like the DSMC method, the DVM works well for highly rarefied gas flows, but encounters great difficulties for near continuum flows~\cite{xu2014direct,chen2015comparative}. Some semi-implicit and implicit DVMs have been developed to remove the
restriction of the time step and improve the efficiency~\cite{yang1995rarefied,pieraccini2007implicit,dimarco2013asymptotic}.

In order to develop an efficient DVM for the whole flow regime, significant effort has been made recently to develop the asymptotic preserving (AP)
schemes~\cite{mieussens2000discrete,pieraccini2007implicit,dimarco2013asymptotic,li2004study,bennoune2008uniformly,filbet2010class,xu2010unified,mieussens2013asymptotic}.
An AP method is uniformly stable with respect to $Kn$, and when $Kn$ is very small, it is consistent with the Chapman-Enskog representation in the continuum limit
~\cite{xu2010unified,mieussens2014survey}.  Therefore, the AP property is critical to a multi-scale method.
Unfortunately, the most AP schemes can only recover the Euler solutions in the hydrodynamic limit, except for the recently developed  unified gas-kinetic
scheme (UGKS)~\cite{xu2014direct,chen2015comparative,xu2010unified,huang2012unified,liu2014unified,chen2016simplification,zhu2016implicit}
and the discrete unified gas-kinetic scheme (DUGKS)~\cite{guo2013discrete,guo2015discrete,wang2015comparative,wang2015coupled,guo2016discrete,wang2016comparison}, which recover the NS solutions.  Both the UGKS and DUGKS
share the same merit that the molecular transport process is coupled with the molecular collision, so that the time step and mesh size are independent of the collision time
and the mean free path, respectively \cite{xu2014direct}.

The main difference between the UGKS and DUGKS lies in the construction of the distribution function across
the cell interface: the UGKS uses the local integration solution of the kinetic model, while the DUGKS adopts its discrete characteristic solution,
thereby avoids computing the complicated gradients of macroscopic variables. Also,
owing to auxiliary functions introduced, the DUGKS only updates single distribution function in the evolution process, while in the UGKS macroscopic variables and distribution
function are updated within a time step. Therefore, the DUGKS is better than  the UGKS in terms of simplicity and efficiency, while their accuracies are at the same level~\cite{zhu2016discrete,wang2015unified}.

So far, the DVM can be roughly classified into two types: the traditional DVM and new AP DVM. The detailed comparison of these two methods will provide essential
information for users to choose the appropriate one for applications.
In this paper, we will perform a comparative study of these two type DVMs in all flow regimes, aiming to clarify their applicability for different flow problems.
It is usually recognized that it is not easy for a second-order accurate traditional DVM to simulate the continuum flow due to the limitations of mesh size and time step, hence
a third-order accurate time-implicit Godunov DVM (GDVM)~\cite{yang1995rarefied} is adopted here in the whole flow regime
including the hydrodynamic regime. On the other hand, it has been demonstrated that the DUGKS, as a newly developed AP DVM, can dynamically describe
flows from the free molecular to hydrodynamic regimes and simultaneously preserve a second-order accuracy in both the spatial and temporal spaces~\cite{wang2015comparative,zhu2015performance,zhu2016discrete}.
Although the two methods are derived from the same model equation, different algorithms will lead to solution discrepancy. In this work, we will analyze these
two typical DVMs in terms of accuracy and efficiency.

The remaining part of this paper is organized as follows. We first make a brief introduction of the time-implicit GDVM and DUGKS, as well as an analysis of
both methods in Sec.~\ref{numerical_methods}. The detailed comparison of these two methods regarding accuracy and efficiency is given in Sec.~\ref{results},
followed by discussions and conclusions in Sec.~\ref{conclusion}.

\section{Numerical methods}

In this section, the GDVM~\cite{yang1995rarefied} and DUGKS~\cite{guo2015discrete} to solve the Shakhov kinetic model equation for monatomic gases~\cite{shakhov1968generalization} is introduced and analyzed.

\label{numerical_methods}

\subsection{The Shakhov model}

In the absence of external force, the Shakhov kinetic model equation can be written as
\begin{equation}
\frac{\partial f}{\partial t} + \bm \xi \cdot \nabla f = - \frac{1}{\tau}\left[ f - f^S \right],
\label{eq:shak}
\end{equation}
where $f = f(\bm x,\bm \xi, t)$ is the velocity distribution function of gas molecules with the molecular velocity $\bm \xi = (\xi_x, \xi_y, \xi_z)$ at the position
$\bm x = (x, y, z)$ and the time $t$, and $f^S$ is the reference equilibrium distribution function expressed by the Maxwellian distribution function $f^{eq}$ and a heat flux correction term:
\begin{equation}
f^S = f^{eq} \left[ 1 + (1-\text{Pr})\frac{\bm c \cdot \bm q}{5pRT}\left( \frac{c^2}{RT} -5) \right) \right ]
= f^{eq} + f_{Pr},
\label{eq:feq_shak}
\end{equation}
where $Pr$ is the Prandtl number, $\bm c = \bm \xi - \bm U$ is the peculiar velocity with $\bm U$ being the macroscopic flow velocity,
$\bm q= \frac{1}{2}\int \bm{c} c^2f\text{d}\bm\xi$ is the heat flux, $R$ is the specific gas constant, and $T$ is the temperature of the gas.
The collision time $\tau$ in Eq.~\eqref{eq:shak} is related to the dynamic viscosity $\mu$ and pressure $p$ by $\tau = \mu/p$.
The Maxwellian distribution function $f^{eq}$ is given by
\begin{equation}
f^{eq} = \frac{\rho}{(2\pi RT)^{3/2}}\exp\left( - \frac{c^2}{2RT} \right),
\label{eq:eq_max}
\end{equation}
where $\rho$ is the gas density.

The conservative variables $\bm W \equiv (\rho, \rho \bm U, \rho E)^T$ are calculated
from the velocity moments of the distribution function:
\begin{equation}
\bm W = \int \bm \psi f \text{d}\bm\xi,
\label{eq:mom}
\end{equation}
where $\bm \psi = \left( 1, \bm \xi, \frac{1}{2}\xi^2 \right )^T$ and
$\rho E = \frac{1}{2}\rho U^2  + \frac{3}{2}\rho RT$ is the total energy.


Since only two-dimensional (2D) problem is considered in this work,  two reduced velocity distribution functions are introduced to cast the three-dimensional molecular velocity space into 2D \cite{yang1995rarefied}:
\begin{subequations}
\begin{align}
g = &\int f(\bm x,\bm \xi, t) \text{d}\xi_z,
\label{eq:g} \\
h = &\int \xi_z^2  f(\bm x, \bm \xi, t) \text{d}\xi_z.
\label{eq:h}
\end{align}
\end{subequations}
For convenience, in what follows we denote $\bm{\xi}=(\xi_x, \xi_y)$ and $\bm{x}=(x,y)$.  Thus, based on $g$ and $h$, we can compute macroscopic variables by
\begin{equation}
\rho = \int g \text{d}\bm \xi,
\quad \rho \bm U = \int \bm \xi g \text{d}\bm \xi,
\quad \rho E = \frac{1}{2}\int (\xi^2 g + h) \text{d} \bm \xi, \quad \bm q = \frac{1}{2}\int \bm c(c^2 g + h) \text{d} \bm \xi.
\label{eq:macro_g}
\end{equation}

The governing equations for the two reduced
distribution functions can be deduced from Eq.~\eqref{eq:shak} as
\begin{subequations}
\begin{align}
\frac{\partial g}{\partial t} + \bm \xi \cdot \nabla g = &\Omega_g = - \frac{1}{\tau}\left[ g - g^S \right],
\label{eq:g_evo} \\
\frac{\partial h}{\partial t} + \bm \xi \cdot \nabla h = &\Omega_h = - \frac{1}{\tau}\left[ h - h^S \right],
\label{eq:h_evo}
\end{align}
\end{subequations}
where the reduced reference distribution functions $g^S$ and $h^S$ are
\begin{subequations}
\begin{align}
g^S(\bm x, \bm \xi, t) = & \int f^S(\bm x,\bm \xi,\xi_z, t) \text{d}\xi_z = g^{eq} + g_\text{Pr},
\label{eq:gs_evo} \\
h^S(\bm x, \bm \xi, t) = & \int \xi_z^2 f^S(\bm x, \bm \xi, \xi_z, t) \text{d}\xi_z= h^{eq} + h_\text{Pr},
\label{eq:hs_evo}
\end{align}
\label{eq:gh_evo}
\end{subequations}
with
\begin{subequations}
\begin{align}
g^{eq} = &\frac{\rho}{ 2\pi RT}\exp\left[ -\frac{c^2}{2RT} \right],
\label{eq:g_eq} \\
h^{eq} = &RTg^{eq},
\label{eq:h_eq} \\
g_\text{Pr} = & (1 - \text{Pr}) \frac{\bm c \cdot \bm q}{5pRT}
\left[ \frac{c^2}{RT}-4\right]g^{eq},
\label{eq:g_pr}\\
h_\text{Pr} = & (1 - \text{Pr}) \frac{\bm c \cdot \bm q}{5pRT}
\left[\frac{c^2}{RT} -2\right]h^{eq}.
\label{eq:h_pr}
\end{align}
\end{subequations}

It is clear that the updating rules for $g$ and $h$ in Eq.~\eqref{eq:gh_evo} have the same structure
\begin{equation}
\frac{\partial \phi}{\partial t} + \bm \xi \cdot \nabla \phi = \Omega = - \frac{1}{\tau}\left[ \phi - \phi^S \right],
\label{eq:phi_evo}
\end{equation}
where the generic symbol $\phi$ is used to denote $g$ or $h$.

Note that the dynamic viscosity $\mu$ for the
hard-sphere (HS) or variable hard-sphere model (VHS) is
\begin{equation}\label{mu_te}
\mu=\mu_{ref}\left(\frac{T}{T_{ref}}\right)^{\omega},
\end{equation}
where $\mu_{ref}$ is the reference viscosity at the reference temperature $T_{ref}$,
$\omega$ is the index related to HS or VHS model, and $\mu_{ref}$ is related to the mean free path $\lambda_{ref}$ as
\begin{equation}\label{viscosity}
\lambda_{ref}=\frac{\mu_{ref}}{p}\sqrt{\frac{\pi R T_{ref}}{2}}.
\end{equation}
By using the Knudsen number $(Kn)$, Mach number $(Ma)$ and Reynolds number $(Re)$, which are respectively defined by
\begin{equation}
Kn=\frac{\lambda_{ref}}{L_{ref}},~ Ma=\frac{U_{ref}}{\sqrt{\gamma R T_{ref}}},~ Re=\frac{\rho_{ref}U_{ref}L_{ref}}{\mu_{ref}},
\end{equation}
the relation Eq.~\eqref{viscosity} leads to
\begin{equation}
Kn=\sqrt{\frac{\pi \gamma}{2}}\frac{Ma}{Re},
\end{equation}
where $\gamma$ is the specific heat ratio, $L_{ref}$, $U_{ref}$ and $\rho_{ref}$ are the reference length, velocity and density, respectively.

\subsection{The traditional discrete velocity method}

The traditional DVM we adopt here is also based on Eq.~\eqref{eq:phi_evo} which is discretized in time by
the fully time-implicit Godunov-type scheme~\cite{yang1995rarefied,titarev2007conservative}:

\begin{equation}
\begin{aligned}\left(\frac{1}{\Delta t^{n}}+{\bm \xi}\cdot \nabla+\frac{1}{\tau^n}\right)\Delta\phi^{n}=\text{RHS}^{n},\\
\text{RHS}^{n}=\frac{1}{\tau^n}\left(\phi^{S, n}-\phi^{n}\right)-{\bm \xi}\cdot \nabla \phi^{n},
\end{aligned}
\label{eq:implicit_Godunov}
\end{equation}
where  $\Delta\phi^{n}=\phi^{n+1}-\phi^{n}$ needs
to be determined at each time step. The right-hand side $\text{RHS}^{n}$ of Eq.~\eqref{eq:implicit_Godunov}
is the explicit part, where the spatial derivative is approximated
by the third-order upwind scheme. In this work, the derivative
with respect to the mesh point $x=x_{j}$ is evaluated by
\begin{equation}\label{third_upwind}
\frac{\partial\phi^{n}}{\partial x}\Biggl|_{j}=\begin{cases}
\frac{2\phi_{j+1}^{n}+3\phi_{j}^{n}-6\phi_{j-1}^{n}+\phi_{j-2}^{n}}{6\Delta x} ,&\: \xi_{x}>0\\
\frac{-2\phi_{j-1}^{n}-3\phi_{j}^{n}+6\phi_{j+1}^{n}-\phi_{j+2}^{n}}{6\Delta x} ,&\: \xi_{x}<0
\end{cases}.
\end{equation}
On the other hand, the left-hand side of Eq.~\eqref{eq:implicit_Godunov} is the implicit
part, where the spatial derivative is approximated by the first-order
upwind scheme. By marching in appropriate direction, \textit{e.g.}
increasing $x$ in the case of $\xi_{x}>0$, the unknown $\Delta\phi^{n}$
can be obtained directly without iterations.

Note that $\Delta t$ in Eq.~\eqref{eq:implicit_Godunov} is a pseudo-time step that is
defined by the Courant-Friedrichs-Lewy (CFL) condition i.e., $\Delta t=\eta\Delta x_{min}/\xi_{max}$,
where $\eta$ is the CFL number, $\Delta x_{min}$ is minimum grid spacing,
and $\xi_{max}$ is the maximum discrete velocity. However, here the CFL number $\eta$
can be smaller than $1$ to capture the transient behavior, it can also be set as large
as $10^{4}$  for steady-state flow problems.

\subsection{Discrete unified gas-kinetic scheme}


The DUGKS is an explicit finite-volume method to solve Eq.~\eqref{eq:phi_evo}.
The computational domain is first divided into some control cells;
then integrating Eq.~\eqref{eq:phi_evo} in a cell $V_j$ (centered at $\bm x_j$)  from time $t_n$ to $t_{n+1}$ $(\Delta t = t_{n+1} -t_n)$,
and using the trapezoidal and middle-point rules for the time integration of collision and convection terms, respectively,
we can obtain evolution equation of DUGKS:
\begin{equation}
\tilde{\phi}_j^{n+1} = \tilde{\phi}_j^{+,n} -\frac{\Delta t}{|V_j|} \mathcal{F}_j^{n+1/2},
\label{eq:phi_tilde_evo}
\end{equation}
where
\begin{subequations}
\begin{align}
\tilde{\phi}& = \phi - \frac{\Delta t}{2}\Omega = \frac{2\tau + \Delta t}{2\tau}\phi - \frac{\Delta t}{2\tau}\phi^S
\label{eq:phi_tilde}, \\
\tilde{\phi}^+& = \phi + \frac{\Delta t}{2}\Omega = \frac{2\tau - \Delta t}{2\tau + \Delta t}\tilde{\phi} + \frac{2 \Delta t}{2\tau + \Delta t}\phi^S,
\label{eq:phi_tilde_plus}
\end{align}
\end{subequations}
are two auxiliary distribution functions, and
\begin{equation}\label{microflux}
 \mathcal{F}^{n+1/2}=\int_{\partial V_j}\left(\bm \xi \cdot \bm n\right)\phi\left(\bm x,\bm \xi, t_{n+1/2}\right)d{\bm S}
\end{equation}
is the micro-flux across cell interface, here $|V_j|$ and $\partial
V_j$ are the volume and surface of cell $V_j$, $\bm{n}$ is the
outward unit vector normal to  cell interface.

Based on the conservative property of collision operators: $
\int{\Omega_g\bm{d\xi}}=0$, $ \int{\bm{\xi}\Omega_g\bm{d\xi}}=\bm{0}$, and $\int{(\xi^2\Omega_g+\Omega_h)\bm{d\xi}}=0$,
we can compute the macroscopic variables from
\begin{equation}
\rho = \int \tilde{g} \text{d}\bm \xi,
\quad \rho \bm U = \int \bm \xi \tilde{g} \text{d} \bm\xi,
\quad \rho E = \frac{1}{2}\int (\xi^2 \tilde{g} + \tilde{h}) \text{d} \bm \xi,
\label{eq:macro_tg}
\end{equation}
and heat flux from
\begin{equation}
\bm q = \frac{2\tau}{2\tau+\Delta t\text{Pr}} \tilde{\bm q}, ~ \text{with} ~
\tilde{\bm q} = \frac{1}{2}\int\bm c(c^2 \tilde g + \tilde h) \text{d}\bm\xi.
\label{q_form_tilde}
\end{equation}
Therefore, in actual implementation, the evolution of $\tilde{\phi}$ is
tracked according to Eq.~\eqref{eq:phi_tilde_evo}, instead of the original distribution functions $\phi$,
to avoid implicit computations.

The key procedure in updating $\tilde{\phi}$ is to evaluate the micro-flux $\mathcal{F}$, which is
solely determined by the gas distribution function $\phi^{n+1/2}(\bm x_f,\bm \xi)$ at the cell
interface $\bm x_f$ and half time step $t_{n+1/2}$. To do so, in the DUGKS Eq.~\eqref{eq:phi_evo}
is integrated along the characteristic line within a half time step $s=\Delta t/2$,
\begin{equation}
\phi ^{n+1/2}(\bm x_f, \bm \xi)- \phi^n( \bm x_f - \bm \xi s, \bm \xi)
= \frac{s}{2}\left[ \Omega^{n+1/2}(\bm x_f, \bm \xi) + \Omega^n (\bm x_f-\bm \xi s, \bm \xi) \right],
\label{eq:phi_half_evo}
\end{equation}
where time integration of the collision term is approximated by the trapezoidal rule. Again, in order to
remove the implicity of Eq.~\eqref{eq:phi_half_evo}, two distribution functions are introduced
\begin{subequations}\label{eq:phi_auxi}
\begin{align}
\bar{\phi} & = \phi - \frac{s}{2}\Omega = \frac{2\tau + s}{2\tau}\phi - \frac{s}{2\tau}\phi^S,
\label{eq:phi_bar}\\
\bar{\phi}^+ & = \phi + \frac{s}{2}\Omega = \frac{2\tau - s}{2\tau + s}\bar{\phi} - \frac{2s}{2\tau + s}\phi^S
\label{eq:phi_bar_plus}.
\end{align}
\end{subequations}
Then Eq.~\eqref{eq:phi_half_evo} is expressed explicitly as
\begin{equation}
\bar{\phi}^{n+1/2}(\bm x_f, \bm \xi) = \bar{\phi}^{+,n}(\bm x_f - \bm \xi s,\bm \xi),
\label{eq:phi_bar_evo_d}
\end{equation}
where $\bar{\phi}^{+,n}$ is constructed as
\begin{equation}
\label{eq:bar-phi+}
\bar{\phi}^{+,n}(\bm{x}_f-\bm{\xi}s,\bm{\xi})=
\bar{\phi}^{+,n}(\x_j,\bxi)+(\x_f-\x_j-\bm{\xi}s)\cdot\bm{\sigma}_j,\quad (\x_f-\bm{\xi}s) \in V_j,
\end{equation}
here $\bm{\sigma}_j$ is the slope of $\bar{\phi}^+$ in cell $j$ which is computed by central difference method
in this study. Note that $\bm{\sigma}_j$ can also be approximated by using some numerical limiters for discontinuous
problems \cite{zhu2016discrete}. Once $\bar{\phi}^{+,n}$ is given, the original distribution function across the cell interface
can be calculated from Eq.~\eqref{eq:phi_auxi}:
\begin{equation}
\label{eq:phi-face}
\phi^{n+1/2}(\x_f, \bxi)=\dfrac{2\tau} {2\tau+s}\bar{\phi}^{n+1/2}(\x_f,\bxi)+\dfrac{s}{2\tau+s }\phi^{S,n+1/2}(\x_f,\bxi).
\end{equation}
where $\phi^{S,n+1/2}(\x_b,\bxi)$ is determined by the conserved variables and heat flux
at cell interface $\bm{x}_b$ and half time step $t_{n+1/2}$, which can be evaluated as
\begin{equation}
\label{eq:W-bar}
\rho=\int{\bar{g} d\bm{\xi}},\quad \rho\bm{u}=\int{\bm{\xi}\bar{g}d\bm{\xi}},\quad \rho E=\dfrac{1}{2}\int{(\xi^2 \bar{g}+\bar{h})d\bm{\xi}},
\end{equation}
and
\begin{equation}
\bm{q}=\frac{2\tau}{2\tau+s \text{Pr}}\bar{\bm q},~ \text{with}~ \bar{\bm q}=\frac{1}{2}\int \bm{c}(c^2\bar{g}+\bar{h})d\bm{\xi}.
\end{equation}
Then the micro-flux can be computed by Eq.~\eqref{eq:phi-face}. Finally  $\tilde{\phi}$ at the cell center can be
updated according to Eq.~\eqref{eq:phi_tilde_evo}. Note that the time step in the DUGKS is solely determined by the CFL condition.

Both GDVM and DUGKS presented above are based on continuous velocity space for convenience.
In actual implementation, the continuous velocity space is discretized into a finite discrete velocity set
$\{{\bm \xi_i}\}$ same as that of the traditional DVM \cite{yang1995rarefied}. For example, in the DUGKS, the distribution functions such as $\tilde{g}$ and $\tilde{h}$ are
defined at these discrete velocity points as $\tilde{g}_i$ and $\tilde{h}_i$. Proper quadrature rule, such as the Newton-Cotes and Gauss-Hermite quadrature,
are adopted to approximate the moments,
\begin{equation}
\rho = \sum_i \varpi_i \tilde{g}_i, \quad \rho\bm U = \sum_i \varpi_i \bm \xi_i \tilde{g}_i, \quad
\rho E = \frac{1}{2}\sum_i \varpi_i \left[ \xi_i^2 \tilde{g}_i + \tilde{h}_i \right],
\label{eq:quadrature}
\end{equation}
where $\varpi_i$ is the weight coefficients for the corresponding quadrature rule.

\subsection{Analysis of the DUGKS and GDVM}
\label{analysis}

Both the DUGKS and GDVM are derived from the same Boltzmann model equation, but different considerations in their algorithm constructions determine their distinctive behaviors in flow simulations.

In the DUGKS, the flux is solely determined by gas molecular distribution functions across the cell interfaces, which is constructed on basis of
the discrete characteristic solution of the kinetic model. Based on Eqs.~\eqref{eq:phi_auxi}, \eqref{eq:phi_bar_evo_d} and \eqref{eq:phi-face},
it can be rewritten as
\begin{equation}
\begin{aligned}
\label{eq:phi-facen}
\phi^{n+1/2}(\x_f, \bxi)&=\dfrac{2\tau} {2\tau+s}\bar{\phi}^{+,n}(\bm x_f - \bm \xi s,\bm \xi)+\dfrac{s}{2\tau+s }\phi^{S,n+1/2}(\x_f,\bxi) \\
&=\frac{2\tau-s}{2\tau+s}\phi^n(\bm x_f - \bm \xi s,\bm \xi)+\frac{s}{2\tau+s}\left[\phi^{S,n}(\bm x_f - \bm \xi s,\bm \xi)+\phi^{S,n+1/2}(\x_f,\bxi)\right].
\end{aligned}
\end{equation}
In the right-hand side of Eq.~\eqref{eq:phi-facen}, the first and second terms represent the kinetic and hydrodynamic contributions, respectively.
It indicates that the molecular transport process is coupled with molecular collisions when evaluating flux across the cell interface.
In the continuum and near continuum flow regions, $\Delta t/ \tau \gg 1$, thus, the flux computed from Eq.~\eqref{eq:phi-facen} is mainly
contributed from the hydrodynamic scale solution; however, in highly rarefied flow regimes, the molecular free transport mechanism will play
an important role due to $\Delta t/ \tau \ll 1$; in the transition regime, the time step $\Delta t$ is comparable to $\tau$, thereby the flux evaluation from
 Eq.~\eqref{eq:phi-facen} takes both the kinetic and hydrodynamic physics into account.
Therefore, with variation of the ratio of $\Delta t/ \tau$, the DUGKS can dynamically describe the flow from free molecular to
hydrodynamic regimes. It also has been demonstrated that with the coupled treatment of
molecular transport and collision processes, the numerical dissipation in DUGKS is
$O(\Delta x^2)+O(\Delta t^2)$~\cite{guo2016discrete}.

On the contrary, in the GDVM, the model equation is directly solved using the implicit finite-difference method, and the convection term is approximated by
the upwind scheme, which means that molecules transport across two grid points freely.
Therefore, for flow regimes in which mesh size is much larger than the mean free path, the use of upwind scheme
is obviously inappropriate, since molecules will physically encounter many collisions when they transport such a long distance in a mesh size scale.
Thus, the GDVM requires much finer mesh to resolve the flow in the near continuum regimes \cite{chen2015comparative}.
Note that the adoption of the third-order upwind approximation~\eqref{third_upwind} for the convection term in the explicit part of Eq.~\eqref{eq:implicit_Godunov}
may improve the GDVM's performance in the continuum and near continuum regimes. It is also noted that this finite-difference DVM computes less
equilibrium state distribution functions than the DUGKS with a finite-volume formulation, thus, with the same CFL number,  the GDVM should be faster than DUGKS for each iteration. In addition, it should be bear in mind that the GDVM becomes an implicit method when using a larger CFL number ($\eta \gg 1$),  and it will lead to fast convergence of the GDVM.

Therefore, the DUGKS may work well in the whole flow regime, and the GDVM is preferable for highly rarefied flows, but may encounter great difficulty
in the continuum and near continuum regimes. It should be noted that although the time step $\Delta t$ in these two methods are both determined by the
CFL condition, for GDVM, $\Delta t$ is a pseudo-time step and has no contribution to the numerical error, thereby the results obtained by the GDVM
with small CFL number and the implicit GDVM with larger CFL number have the same accuracy. The above points will be verified in the following simulations.

\section{Numerical results and discussions}
\label{results}
\subsection{Force-driven Poiseuille flow}

\begin{figure}[!htb]
	\centering
	\begin{subfigure}[b]{0.5\textwidth}
		\includegraphics[width=\textwidth]{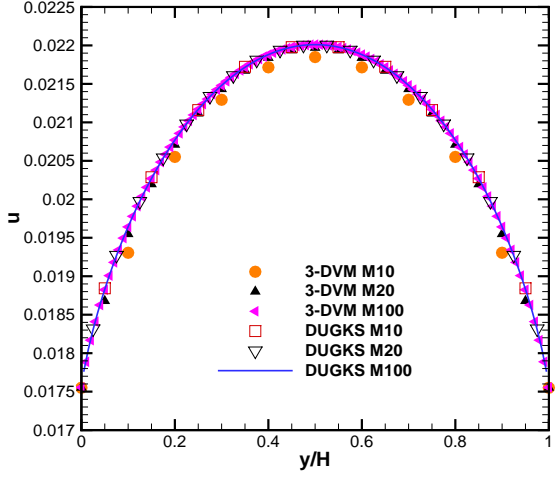}
		\caption{$Kn=10$}
		\label{subfig:vel_kn10}
	\end{subfigure}~
	\begin{subfigure}[b]{0.5\textwidth}
		\includegraphics[width=\textwidth]{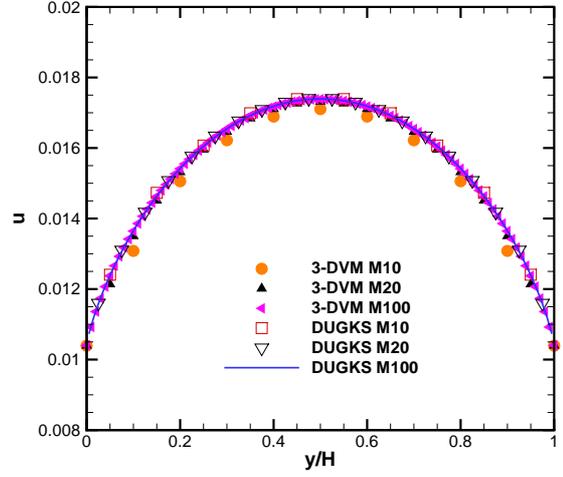}
		\caption{$Kn=1$}
		\label{subfig:vel_kn1}
	\end{subfigure}\\
	\begin{subfigure}[b]{0.5\textwidth}
		\includegraphics[width=\textwidth]{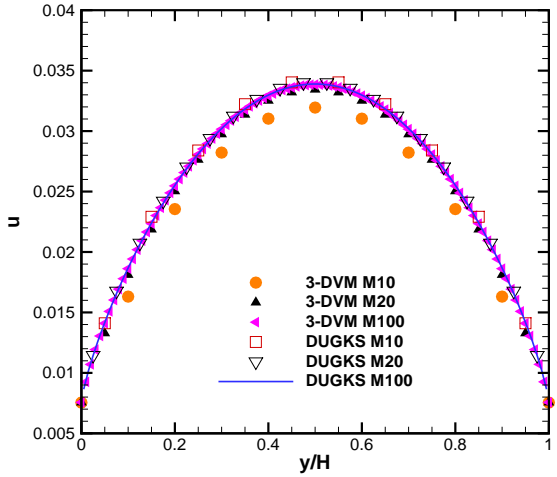}
		\caption{$Kn=0.1$}
		\label{subfig:vel_kn01}
	\end{subfigure}~
	\begin{subfigure}[b]{0.5\textwidth}
		\includegraphics[width=\textwidth]{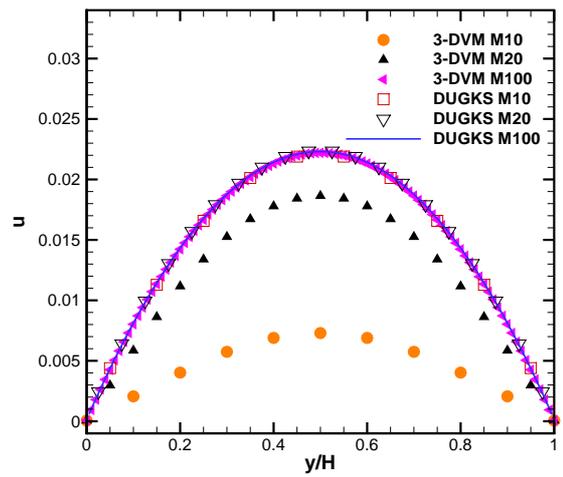}
		\caption{$Kn=10^{-3}$}
		\label{subfig:vel-kn0001}
	\end{subfigure}\\
	\caption{The velocity profiles (normalized by $\xi_0$) along the channel cross-section at  ({a}) $Kn=10$,  $G=0.01$  ({b}) $Kn=1$, $G=0.01$
		({c}) $Kn=0.1$,  $G=0.01$ and ({d}) $Kn=10^{-3}$, $G=10^{-4} $ obtained from the DUGKS and GDVM with different spatial discretizations. M10, M20 and M100 represent the results with $10$, $20$ and $100$ grid points along the channel cross section, respectively,
		and 3-DVM denotes the third-order time implicit GDVM. The same notations are also used in the following figures.}
	\label{fig:vel_poie}
\end{figure}

The performance of the GDVM and DUGKS is first evaluated by simulating the one-dimensional (1D) force-driven Poiseuille flow between two parallel plates,
which are located at $y=0$ and $y=H$. An external force is applied in the $x$-direction, so that the Shakhov model~\eqref{eq:phi_evo} becomes
\begin{equation}\label{SK_force}
\frac{\partial \phi}{\partial t} + \xi_y  \frac{\partial \phi} {\partial y} = \Omega+F_x,
\end{equation}
where $F_x$ is the force term. Suppose the magnitude of the external acceleration $G$ is very small, the force term  can be approximated by
\begin{equation}
F_x=-{G}\frac{\partial \phi}{\partial \xi_x}\approx-{G}\frac{\partial \phi^{eq}}{\partial \xi_x},
\end{equation}
see the expression for $\phi^{eq}$ in Eqs.~\eqref{eq:g_eq} and~\eqref{eq:h_eq}.

In the GDVM, Eq.~\eqref{SK_force} is directly solved by considering $F$ as a source term, while in the DUGKS, the Strang splitting method is used \cite{strang1968construction}:
at the beginning of each time step, the distribution function $\tilde{\phi}^n$ is updated within a half time step by $\partial_t \tilde{\phi}=\Delta t F_x /2$,
and then the procedure of DUGKS is executed followed by updating  $\tilde{\phi}^{n+1}$ within a half time step by using the same way as that at the beginning of each iteration.

In our simulations, we use $10$, $20$, and $100$ mesh points between two parallel plates with the distance $H=1$. The gas flow from the highly rarefied to
the hydrodynamic regimes (the Knudsen number from 10 to $10^{-4}$) is simulated by varying the gas pressure. 
The diffuse boundary condition is
applied on both plates. The hard-sphere gas is considered, where
the exponent $\omega$ in Eq.~\eqref{mu_te} is 0.5.
Our simulations start from a global equilibrium state. The convergence criterion for the steady-state is defined by
\begin{equation}\label{work_state}
E(t)=\frac{\sum \left|\bm{u}(t)-\bm{u}(t-100\Delta t)\right|}{\sum |\bm{u}(t)|}<10^{-6}\\.
\end{equation}

The discretization of the molecular velocity space depends on the rarefaction level of the gas flow.
For all the Knudsen numbers, the continuous molecular velocity space is truncated into a finite range of $[-4, 4]\times[-4, 4]$.
For the cases of $1\leq Kn \leq 10$ and $0.1\leq Kn<1$,  we use the $100\times100$  and $50\times50$  non-uniform discrete velocity points~\cite{wu2014solving},
respectively, while for the cases of $ 0.01 \leq Kn<0.1$ and $ 10^{-4} \leq Kn<0.01$,  $28\times28$ and $8\times8$ half-range
Gauss-Hermit discrete velocity points are applied, respectively. Note that all the parameters presented in this paper are dimensionless,
where the spatial length and molecular velocity are scaled by $H$ and $\xi_{0}=\sqrt{2RT}$.

\begin{figure}[t]
	\centering
	\begin{subfigure}[b]{0.5\textwidth}
		\includegraphics[width=\textwidth]{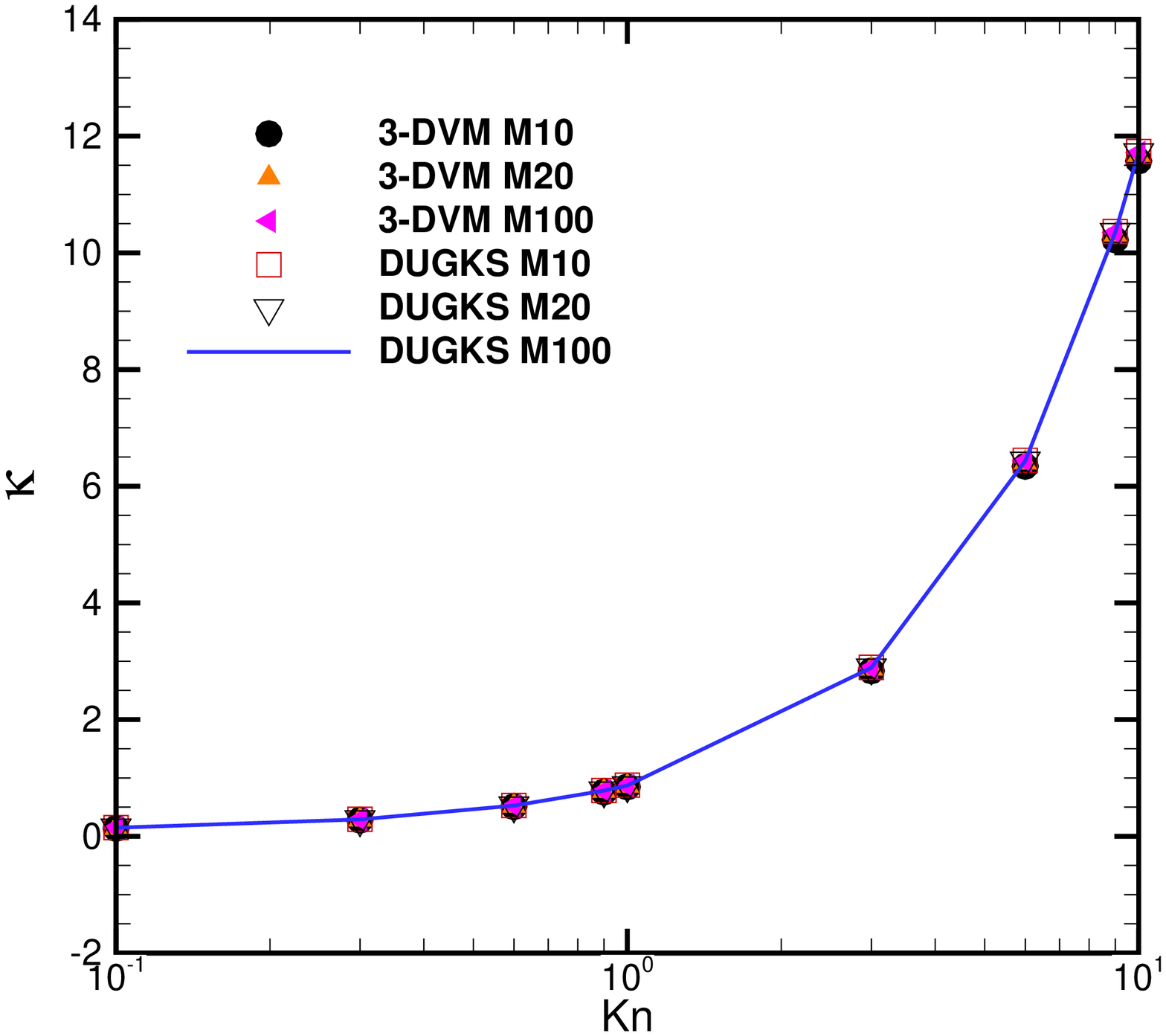}
		\caption{$0.1 \leq Kn \leq 10$}
		\label{subfig:perm_kn10}
	\end{subfigure}~
	\begin{subfigure}[b]{0.5\textwidth}
		\includegraphics[width=\textwidth]{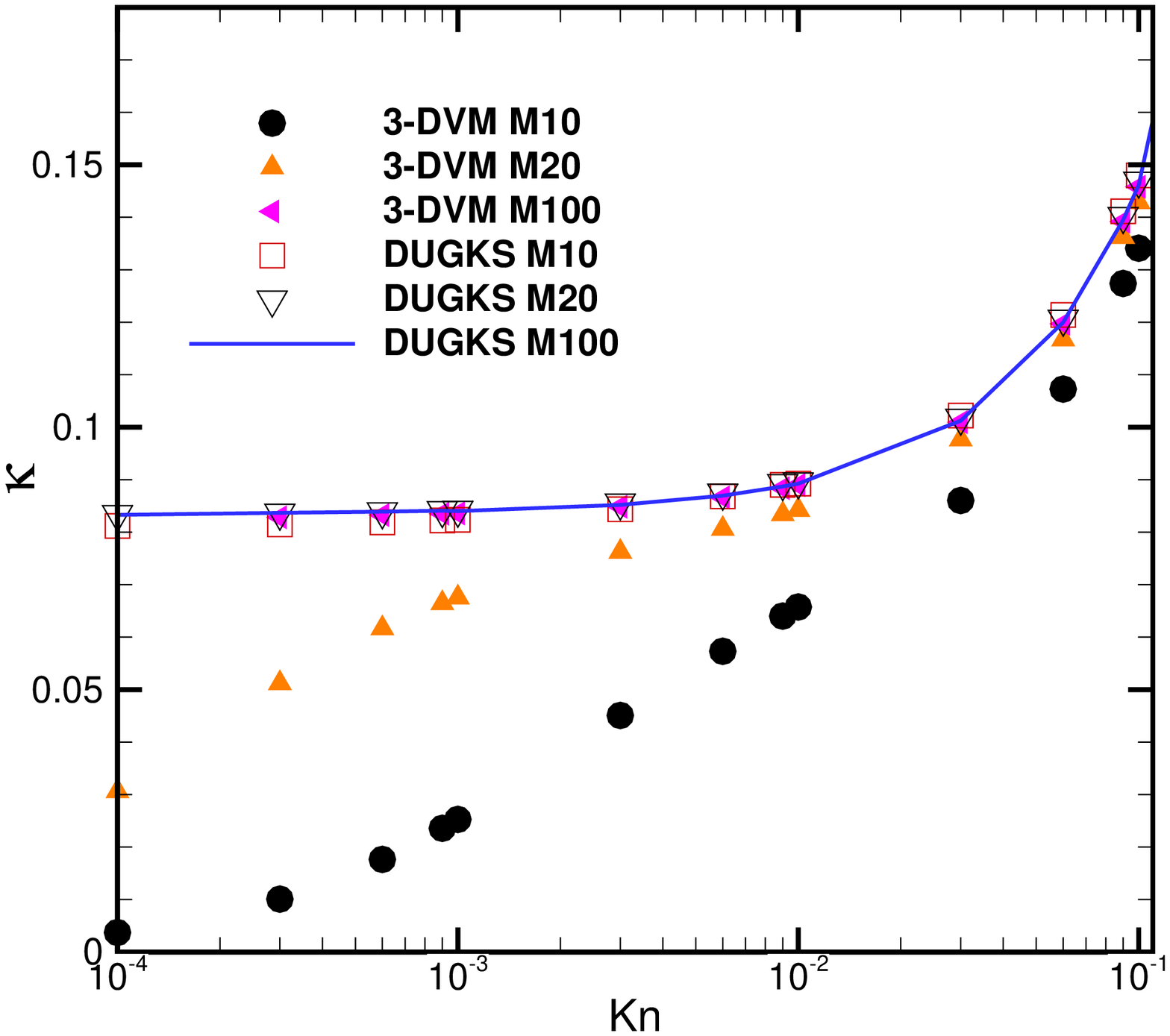}
		\caption{$10^{-4} \leq Kn \leq 0.1$}
		\label{subfig:perm_kn01}
	\end{subfigure}~
	\caption{
		Apparent gas permeability (normalized by $GH^2$) in the Poiseuille flow at different $Kn$: (\textit{a}) $0.1 \leq Kn \leq 10$,
		(\textit{b}) $10^{-4} \leq Kn \leq 0.1$.}
	\label{fig:poi_perm}
\end{figure}

The velocity profiles along the channel cross section at $Kn=10$, 1, 0.1, and $10^{-3}$ are plotted in Fig.~\ref{fig:vel_poie}.
The numerical results of the DUGKS with grid points of $100$ can be regarded as the reference solutions. It is found that the DUGKS can give
adequately accurate results with just 10 grid points in the whole flow regions, while for the GDVM, 20 and 100 mesh points are respectively required in highly rarefied and near continuum regimes.
For example, when $Kn=10^{-3}$,  the GDVM with $20$ mesh points underpredicts the velocity in the channel center  by $16\%$, while that of the DUGKS is only $2\%$ even with a coarser mesh of $10$, see Fig.~\ref{fig:vel_poie}(d).

We then compare the apparent gas permeability $\kappa$  predicted by the GDVM and DUGKS,  which is defined by
\begin{equation}
\kappa= \frac{ 2Kn}{\sqrt{\pi}GH^2} \int_0^H{u(y)}dy.
\end{equation}
Figure~\ref{fig:poi_perm}  shows the permeability  at different Knudsen numbers.
In the free molecular and transition regions (Fig.~\ref{subfig:perm_kn10}), the results obtained from the GDVM and DUGKS agree well with each other at the given meshes.
However, when the flow approaches the slip and continuum regimes (Fig.~\ref{subfig:perm_kn01}),
in order to obtain accurate results, the GDVM requires the spatial mesh that is about one order of magnitude finer than that of the DUGKS.
Note that when $Kn=10^{-4}$, the permeability obtained from the GDVM with 100 mesh points is not presented in Fig.~\ref{subfig:perm_kn01}, since
it is time consuming to reach the steady state.

\begin{figure}[t]
	\centering
	\includegraphics[width=0.5\textwidth]{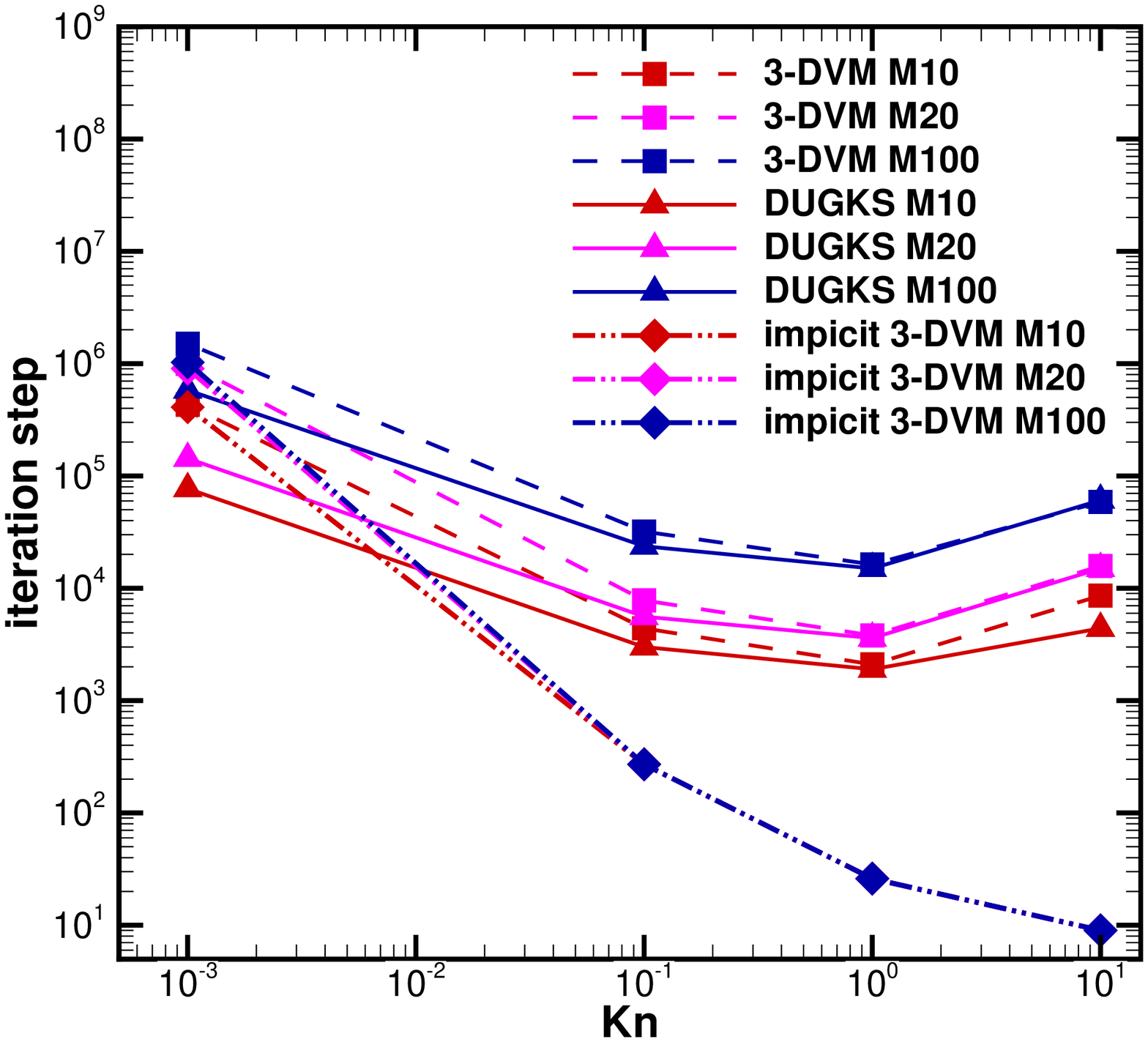}
	\caption{Iteration steps required to attach the stead-state defined by Eq.~\eqref{work_state} at different $Kn$ and meshes. Note that the convergence criteria for implicit GDVM is estimated for one time step.}
	\label{fig:interations}
\end{figure}

\begin{table}[t]
\centering
\caption{\label{tablec} The total CPU time costs (in second) of the implicit GDVM and DUGKS when the results are in reasonable accuracy. The   convergency criteria for implicit GDVM is measured at one time step. }
\begin{tabular}{c c c c c c c c c}
  \hline
   \hline
   $Kn$      & $0.001$ & $0.1$ & $1$ & $10$   \\
  \hline
   \hline
  $t_{DUGKS}$                   & 12.61    & 17.39     & 43.22 & 183  \\
$t_{implicit~GDVM}$             &1485   & 3.79     & 3.53   & 1.25     \\
$t_{DUGKS}/t_{implicit~GDVM}$   & 0.0068    & 4.59    & 12.24 & 146.4  \\
\hline
\hline
\end{tabular}
\end{table}
The above comparisons demonstrates the superiority of the DUGKS over the GDVM in terms of the mesh requirement. However, the computational efficiency is another
important issue. To this end, we study the time needed for each iteration, as well as the iteration numbers needed to reach the convergence.
The CPU time cost for each iteration is  assessed when both codes are executed on the same
workstation (Dual Intel Xeon CPU E5-2630 v3 @ 2.4 GHz with 64Gb of RAM memory).
It is found that for the case of $Kn=10$ with $100$ mesh points, the DUGKS needs $0.0593s$ for each iteration, which is about twice as much as the GDVM. Iteration steps of the GDVM and DUGKS to achieve
the steady-state defined by Eq.~\eqref{work_state} are also given in Fig.~\ref{fig:interations}. With the same CFL number $\eta=0.5$, both methods have the similar
convergency rate in the highly rarefied regimes, while in the near continuum regime, the DUGKS convergents much faster than the GDVM. When using a larger CFL number up to $10^{6}$,
the convergence rate of the implicit GDVM turns to be about two orders of magnitude faster than the explicit DUGKS when $Kn>1$, however,
although using such large CFL number, the GDVM is still about one order of magnitude slower than the DUGKS in the near continuum regime, i.e. $Kn<0.001$.

Moreover, as shown in Figs.~\ref{fig:vel_poie} and \ref{fig:poi_perm},
in order to get the similar accurate results, for the cases of $Kn\leq 0.1$ and $Kn\geq 1$, the GDVM needs respective $100$ and $20$ grid points, while the
DUGKS only requires $10$ mesh points in the whole regime. It means that aiming to get reasonably accurate results, the DUGKS requires fewer mesh points than
the GDVM. As a result, the efficiency of DUGKS may be significantly improved. The total CPU time costs of the implicit GDVM and DUGKS to get results in reasonable accuracy are compiled in table~\ref{tablec}. It is found that the DUGKS is about two orders of magnitude faster than the implicit GDVM in near continuum regime,
while as $Kn$ increases, the implicit GDVM turn to be about two orders of magnitude faster than the DUGKS in the highly rarefied regimes.

\subsection{Lid-driven cavity flow}

In addition to the force-driven Poiseuille flow, the comparative study between the GDVM and DUGKS is
also performed on a 2D lid-driven cavity flow, which is a standard benchmark problem to validate numerical accuracy and efficiency~\cite{chen2015comparative,wang2015comparative,john2010investigation,john2011effects,huang2013unified,guo2013discrete}.
Here the Knudsen number is chosen to be $ Kn=10$,  $1$, $0.1$, $0.0259$, and $6.47\times 10^{-4}$,  so that the flows vary from the free molecular
to hydrodynamic regimes. For the cases of $ Kn=0.0259$ and $6.47\times 10^{-4}$, the corresponding Reynolds numbers are $Re=10$ and $400$, respectively.
The length and height of the cavity are both set to be $1$. The Mach number defined by the velocity of top-wall $U_w$ is 0.16, while the
other three walls are stationary. The temperature of all walls is fixed at $T_w=1$, and the diffuse boundary condition~\cite{guo2013discrete} is used.

\begin{figure}[!htb]
\centering
\begin{subfigure}[b]{0.5\textwidth}
\includegraphics[width=\textwidth]{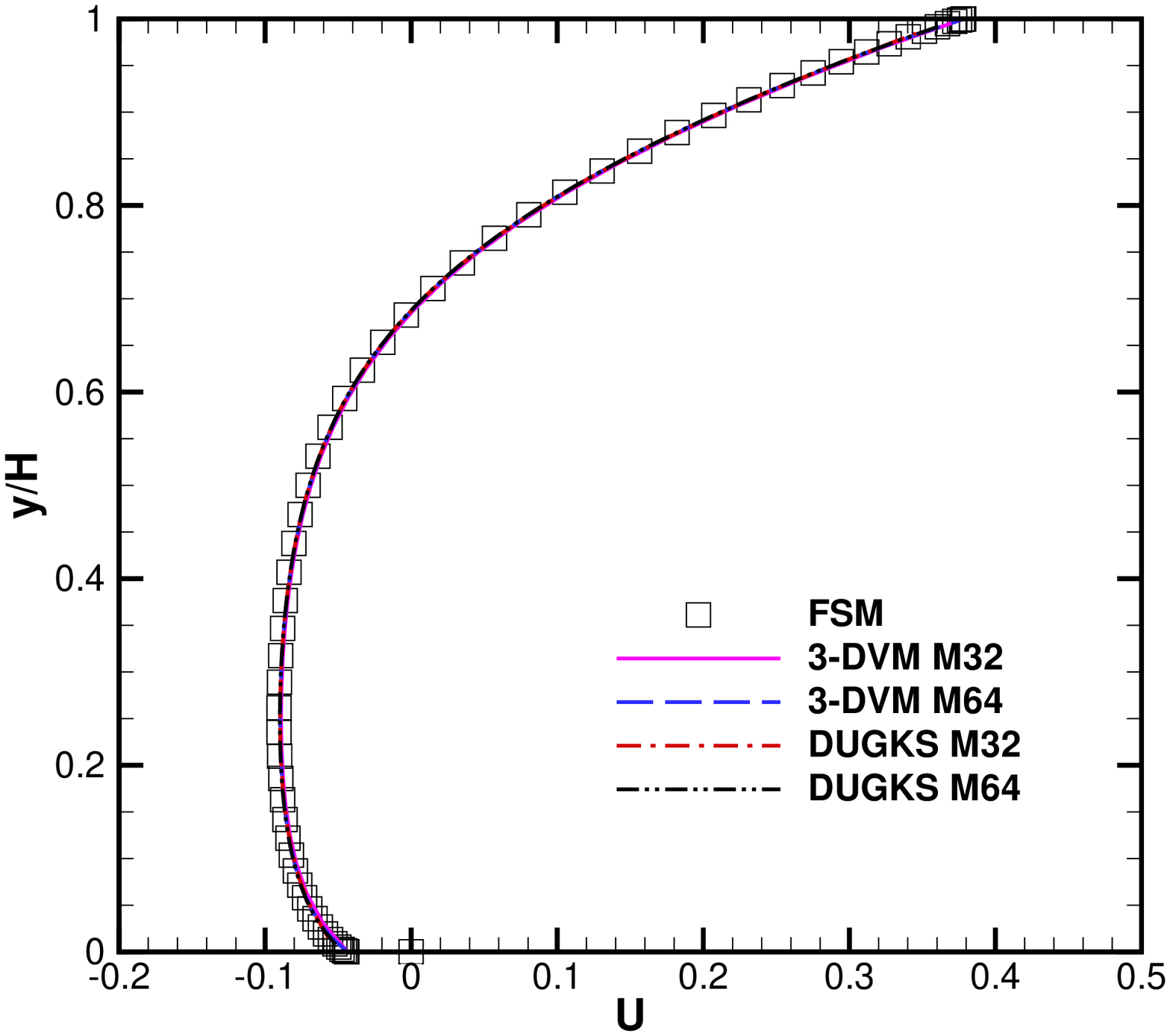}
\caption{$U$}
\label{subfig:kn10_u}
\end{subfigure}~
\begin{subfigure}[b]{0.5\textwidth}
\includegraphics[width=\textwidth]{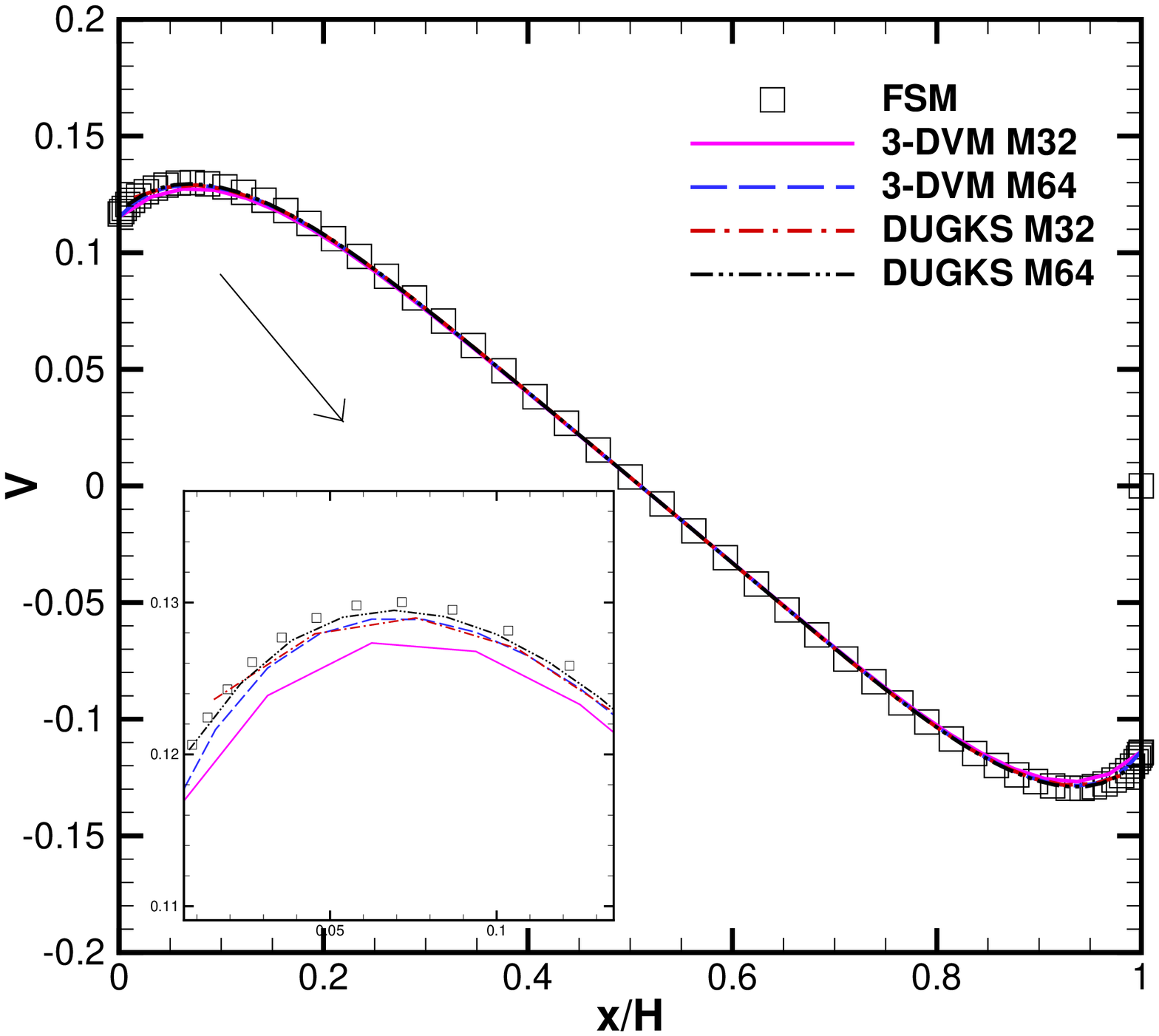}
\caption{$V$}
\label{subfig:kn10_v}
\end{subfigure}\\
\begin{subfigure}[b]{0.5\textwidth}
\includegraphics[width=\textwidth]{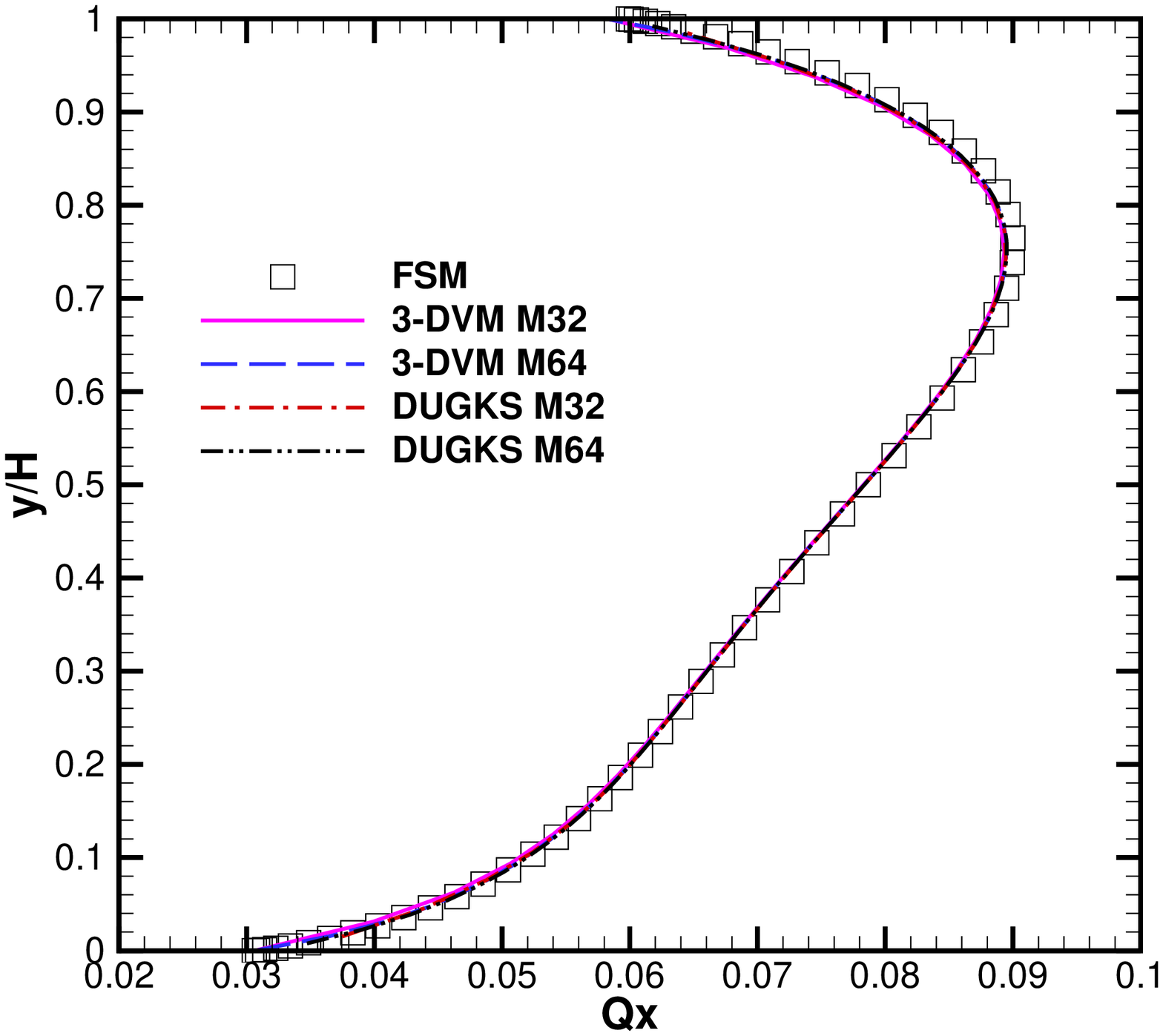}
\caption{$Qx$}
\label{subfig:kn10_qx}
\end{subfigure}~
\begin{subfigure}[b]{0.5\textwidth}
\includegraphics[width=\textwidth]{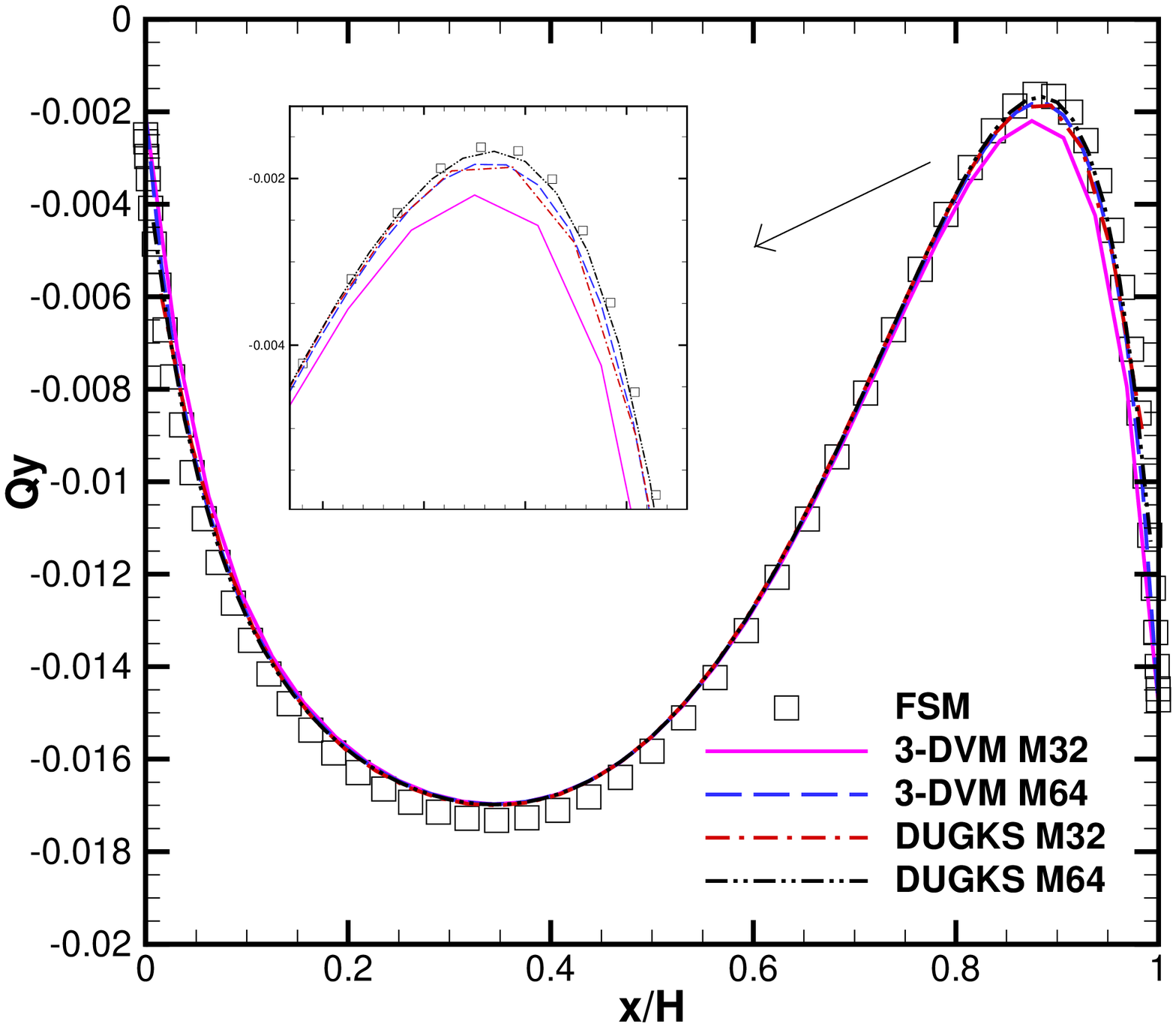}
\caption{$Qy$}
\label{subfig:kn10_qy}
\end{subfigure}\\
\caption{
The results of the cavity flow at $Kn=10$: (\textit{a}) $U$-velocity along the vertical centerline,
 (\textit{b}) $V$-velocity along the horizontal centerline,
 (\textit{c}) heat flux $Qx$ along the vertical centerline
 and (\textit{d}) heat flux $Qy$ along the horizontal centerline.} \label{fig:cavity_kn10}
\end{figure}

In the simulations, the continue molecular velocity space is
discretized in a finite range $[-4, 4] \times [-4, 4] $.
When $Kn=10$ and $1$, we use $100\times 100$ non-uniform discrete velocity points~\cite{wu2014solving}, while
when $Kn=0.1, 0.0259$ and $6.47\times 10^{-4}$, we  apply $28\times28$, $8 \times 8$ and $4\times 4$ half-range
Gauss-Hermit discrete velocity points, respectively. Independence of results with respect to the number of discrete velocity is already validated.
The CFL number $\eta$ in both methods are set to be 0.5 unless otherwise stated. It should be noted that in what follows the ``resolved"  result
means the solution is mesh independent; velocity and heat flux presented are normalized
by $U_w$ and $p_0U_w$, respectively, where $p_0$ is the initial pressure.

\begin{figure}[!htb]
\centering
\begin{subfigure}[b]{0.5\textwidth}
\includegraphics[width=\textwidth]{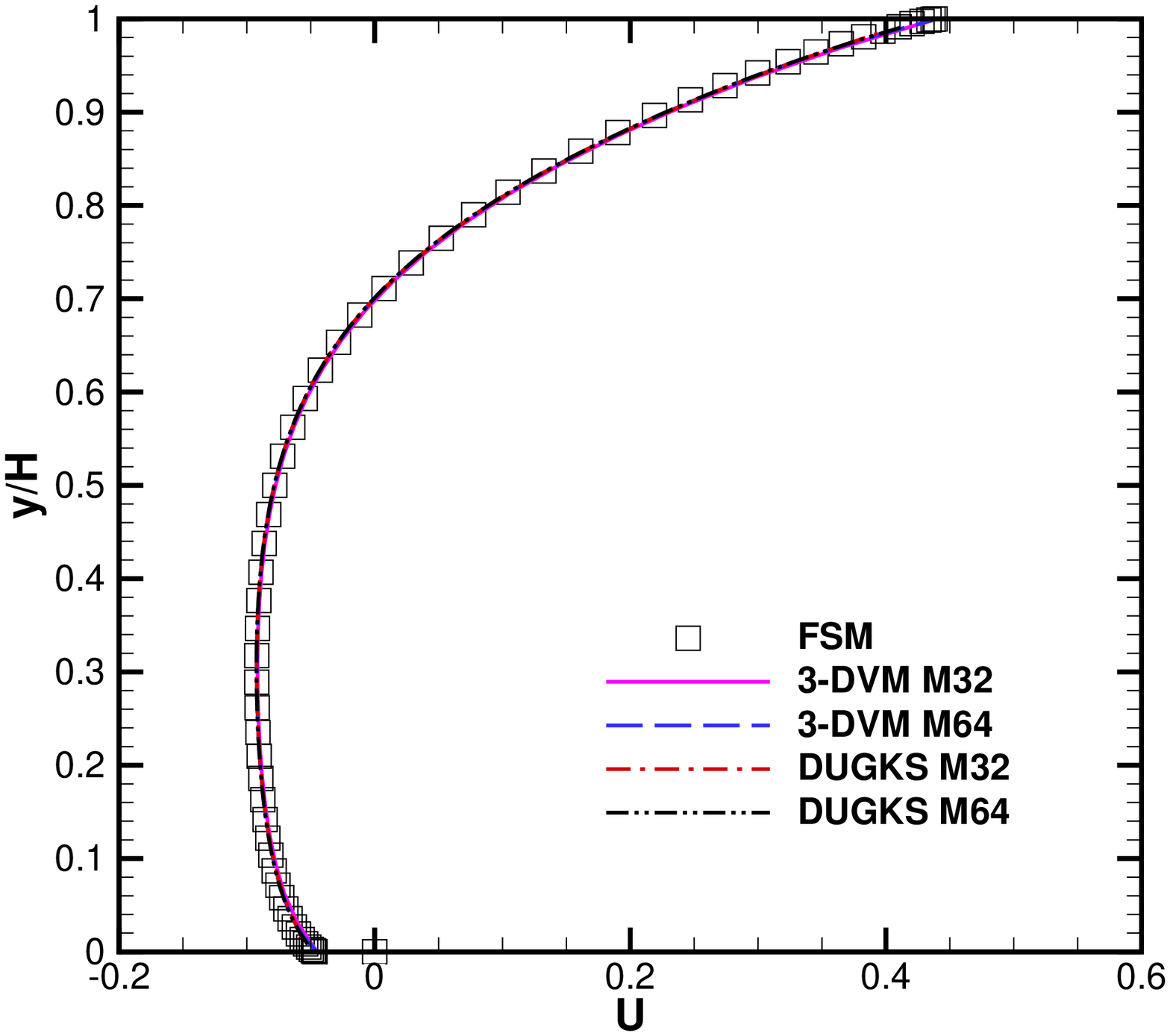}
\caption{$U$}
\label{subfig:kn1_u}
\end{subfigure}~
\begin{subfigure}[b]{0.5\textwidth}
\includegraphics[width=\textwidth]{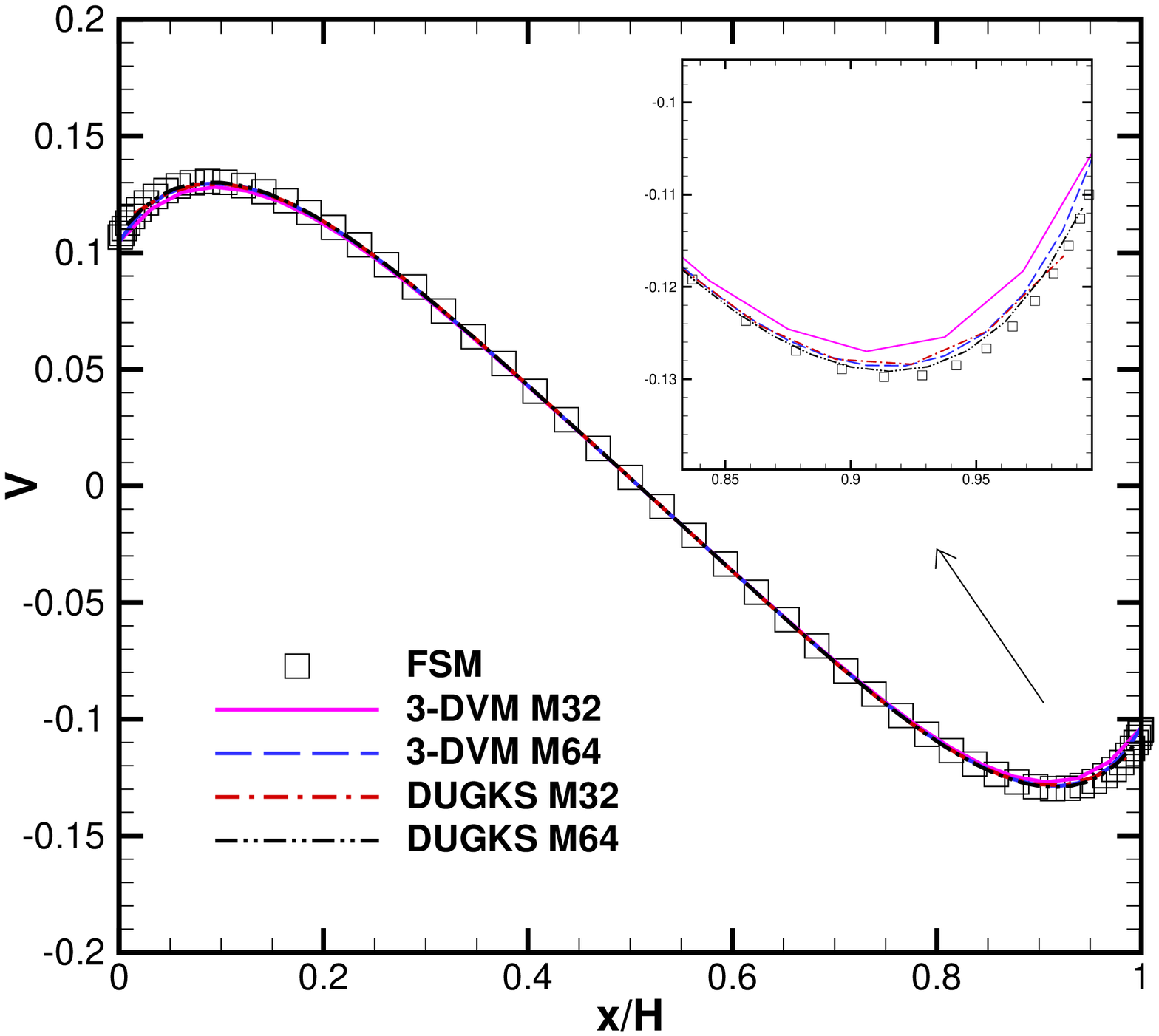}
\caption{$V$}
\label{subfig:kn1_v}
\end{subfigure}\\
\begin{subfigure}[b]{0.5\textwidth}
\includegraphics[width=\textwidth]{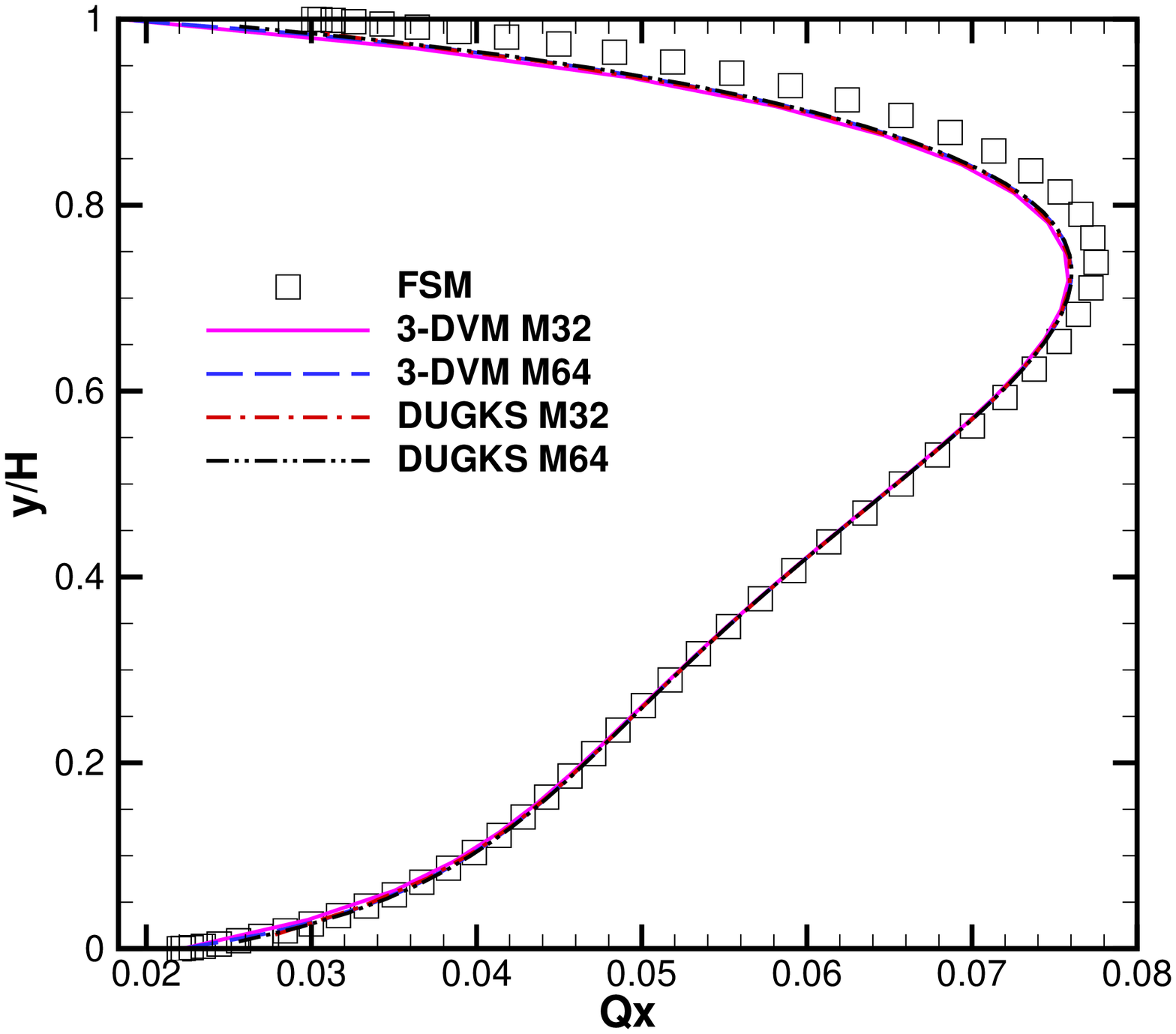}
\caption{$Qx$}
\label{subfig:kn1_qx}
\end{subfigure}~
\begin{subfigure}[b]{0.5\textwidth}
\includegraphics[width=\textwidth]{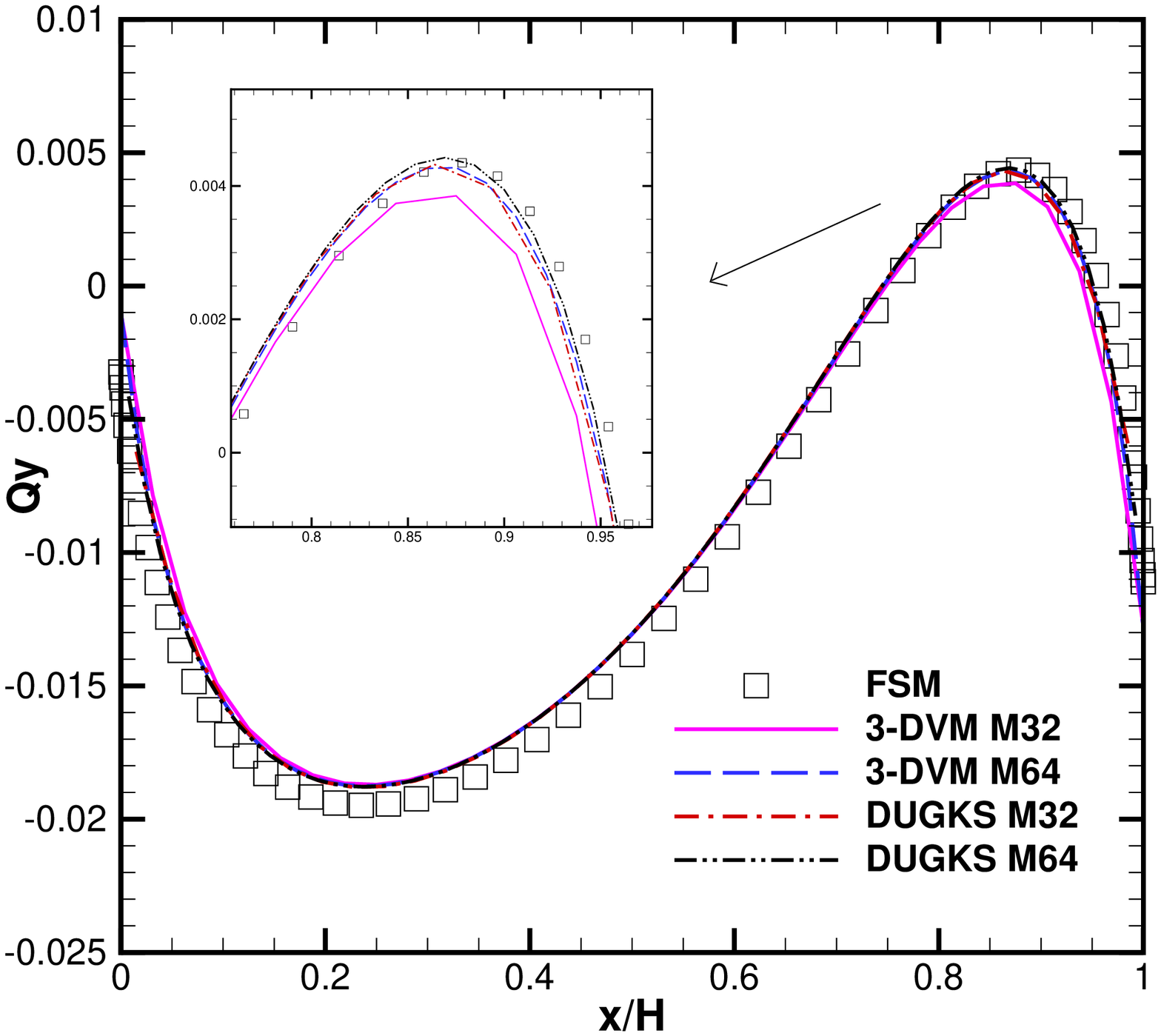}
\caption{$Qy$}
\label{subfig:kn1_qy}
\end{subfigure}\\
\caption{
The results of the cavity flow at $Kn=1$: (\textit{a}) $U$-velocity along the vertical centerline,  (\textit{b}) $V$-velocity along the horizontal centerline,
 (\textit{c}) heat flux $Qx$ along the vertical centerline and (\textit{d}) heat flux $Qy$ along the horizontal centerline.}
\label{fig:cavity_kn1}
\end{figure}

\begin{figure}[!htb]
\centering
\begin{subfigure}[b]{0.5\textwidth}
\includegraphics[width=\textwidth]{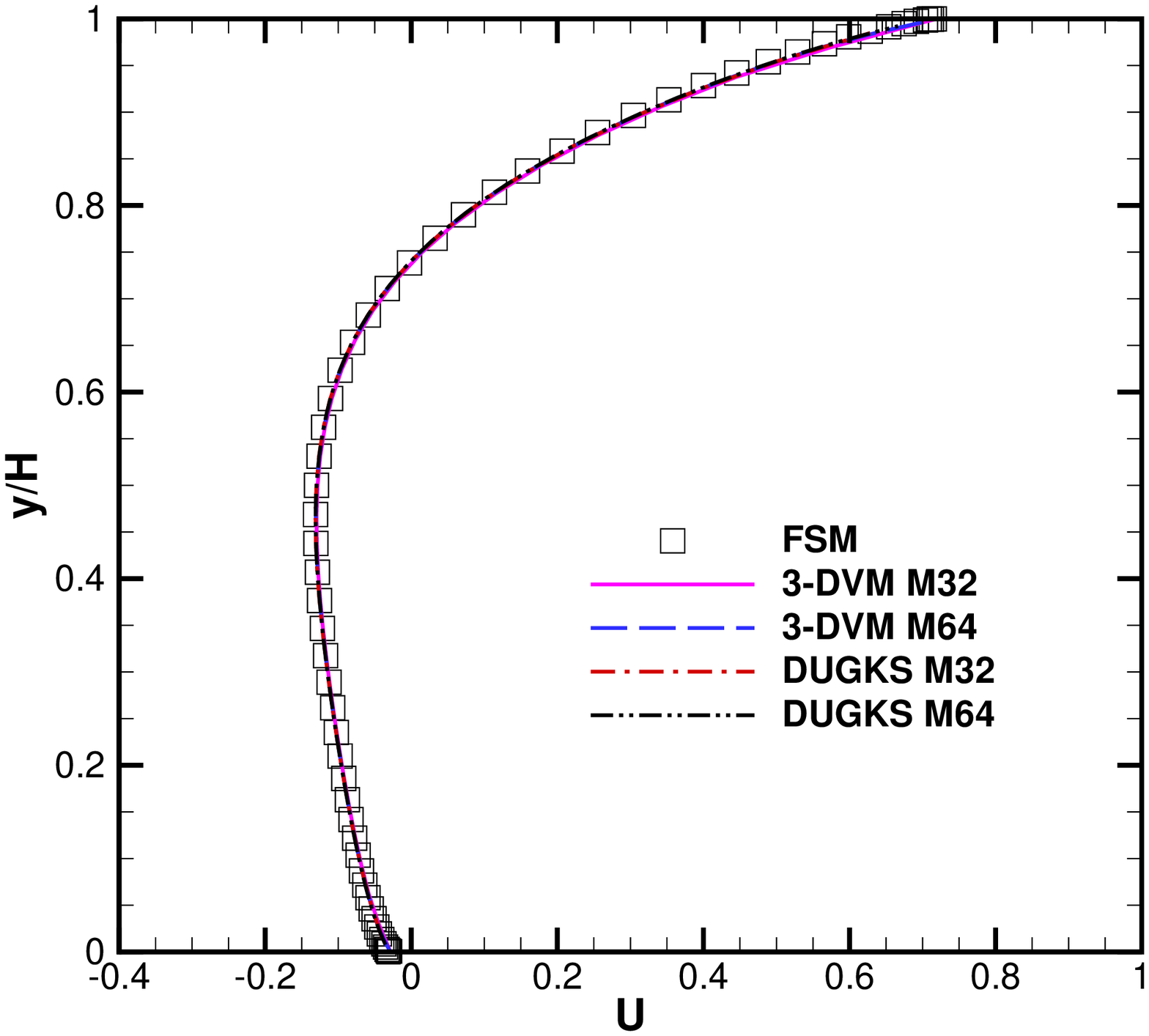}
\caption{$U$}
\label{subfig:kn01_u}
\end{subfigure}~
\begin{subfigure}[b]{0.5\textwidth}
\includegraphics[width=\textwidth]{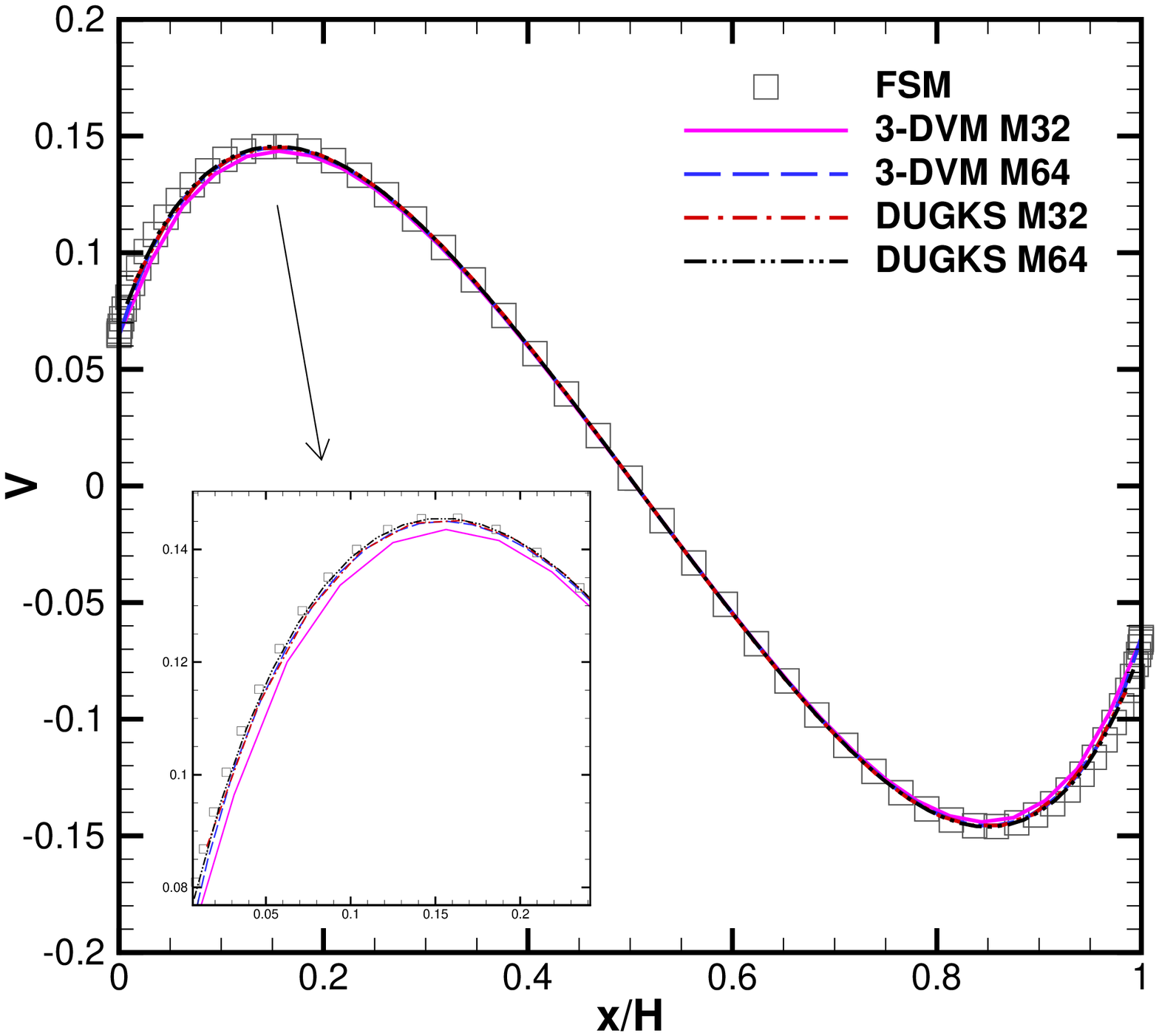}
\caption{$V$}
\label{subfig:kn01_v}
\end{subfigure}\\
\begin{subfigure}[b]{0.5\textwidth}
\includegraphics[width=\textwidth]{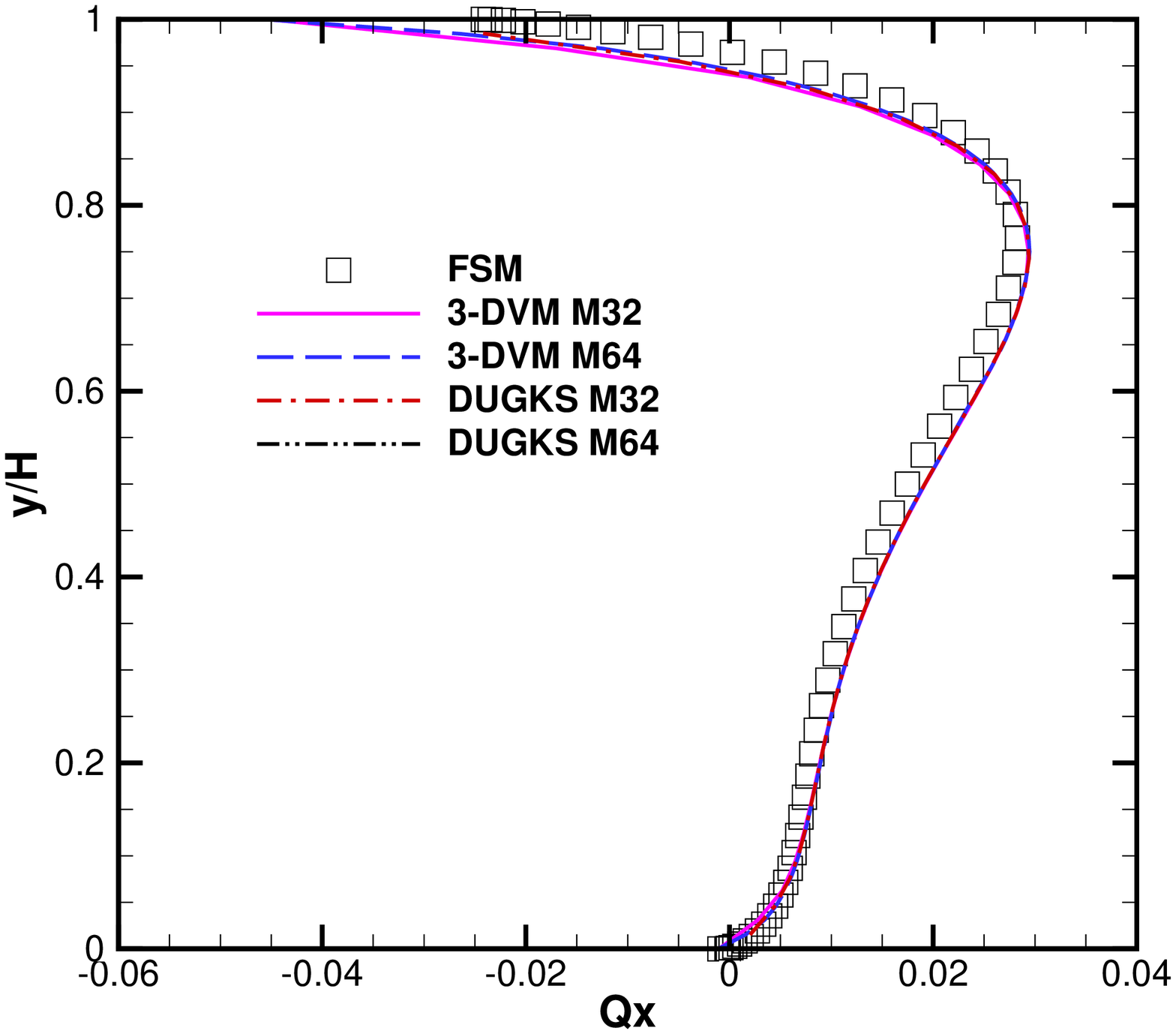}
\caption{$Qx$}
\label{subfig:kn01_qx}
\end{subfigure}~
\begin{subfigure}[b]{0.5\textwidth}
\includegraphics[width=\textwidth]{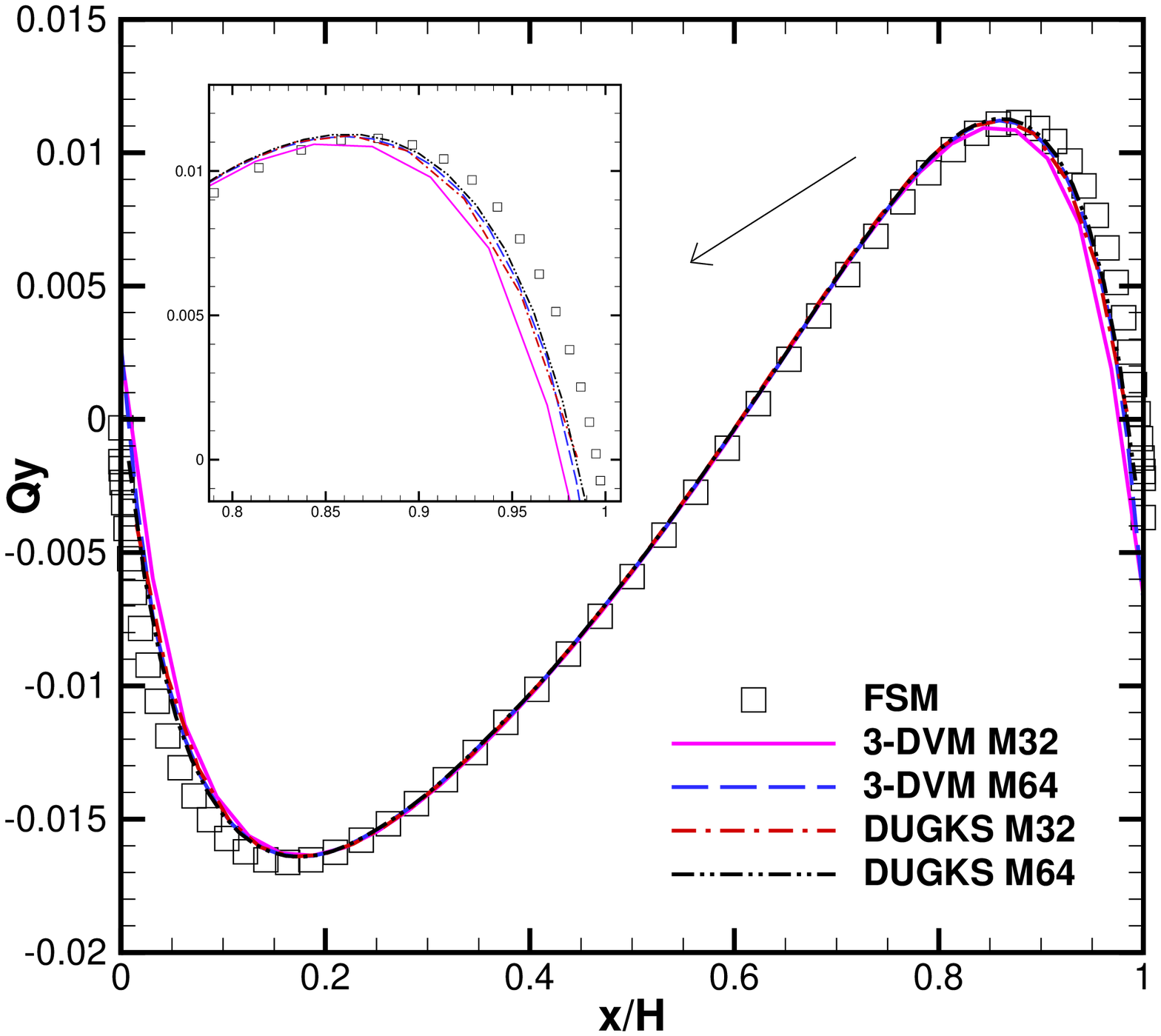}
\caption{$Qy$}
\label{subfig:kn01_qy}
\end{subfigure}\\
\caption{
 The results of the cavity flow at $Kn=0.1$: (\textit{a}) $U$-velocity along the vertical centerline,  (\textit{b}) $V$-velocity along the horizontal centerline,
 (\textit{c}) heat flux $Qx$ along the vertical centerline  and (\textit{d}) heat flux $Qy$ along the horizontal centerline.}
\label{fig:cavity_kn01}
\end{figure}

Figures~\ref{fig:cavity_kn10}, \ref{fig:cavity_kn1} and \ref{fig:cavity_kn01} show the velocity and heat flux profiles along
the horizontal and vertical centerlines of the cavity when $Kn=10,1$ and $0.1$, respectively.
In order to compare the accuracy of these two methods, the results on different mesh resolutions are presented, and with a mesh of
$64^2$ the results are already well-resolved.
The results of the full Boltzmann equation solved by fast spectral method (FSM) are also included for comparison~\cite{wu2013deterministic, wu2014solving, wu2015kinetic}.
As we can see, the resolved results of velocity profiles agree well with those from the FSM, however, discrepancies are observed for heat flux, despite that the resolved results of the GDVM and DUGKS agree well with each other. This can be attributed to that the GDVM and DUGKS are developed based on the simplified
Boltzmann model equation, while the FSM method solves the full Boltzmann model. In addition, the heat flux, a high-order moment
quantity, is more sensitive to the collision model than low-order ones.

In addition, as shown in Figs.~\ref{subfig:kn10_qy}, \ref{subfig:kn1_qy} and \ref{subfig:kn01_qy},
the GDVM with $32 \times 32$ grid points underestimates the peak value of vertical heat flux $Qy$ adjacent to the right wall,
while the DUGKS results with the same coarse mesh are in reasonable agreement with that of the fine mesh of $64^2$.
For instance, for the case of GDVM at $Kn=10$ with mesh of $32 \times 32$, the maximum relative error of $Qy$ is about $38.2\%$ compared with
the resolved results, while it is about $11.1\%$ for the DUGKS counterpart. Additionally, it is interesting to note that
there is no such clear discrepancy for the horizontal heat flux $Qx$. It is because that
the variation of $Qy$ along the horizontal direction is more intensive than that of $Qx$ along the vertical direction.
With such coarse mesh in non-smooth region, the third-order accurate upwind scheme in which the numerical stencil
expands to large distance, may produce large error. The second-order accurate upwind scheme is also tested with coarse mesh of $32^2$
and it captures this $Qy$ peak much better than the high order one with the same mesh, but the results of the high-order GDVM is still overall
better than those of the second-order one.

\begin{figure}[!htb]
\centering
\begin{subfigure}[b]{0.5\textwidth}
\includegraphics[width=\textwidth]{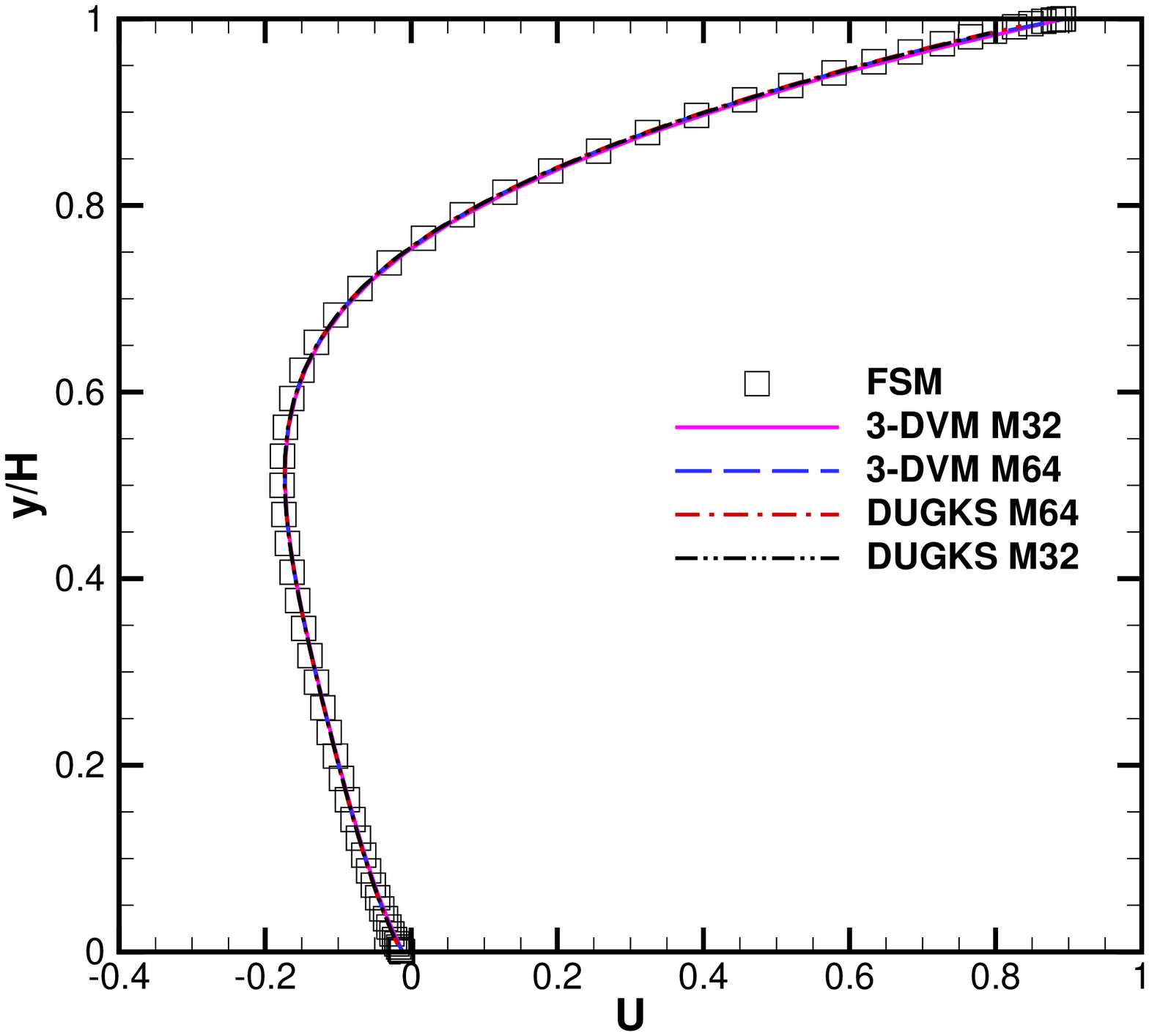}
\caption{$U$}
\label{subfig:re10_u}
\end{subfigure}~
\begin{subfigure}[b]{0.5\textwidth}
\includegraphics[width=\textwidth]{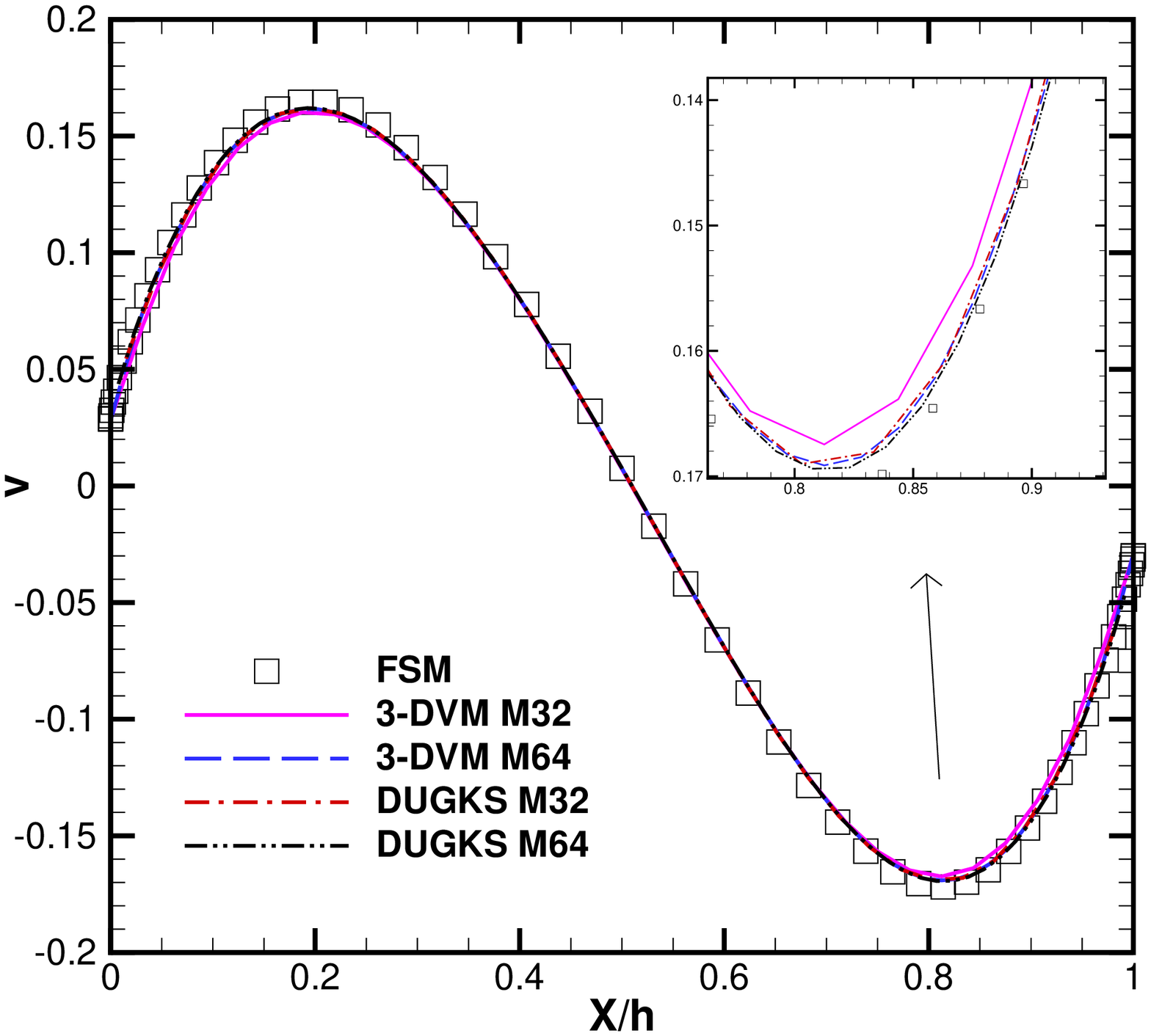}
\caption{$V$}
\label{subfig:re10_v}
\end{subfigure}\\
\begin{subfigure}[b]{0.5\textwidth}
\includegraphics[width=\textwidth]{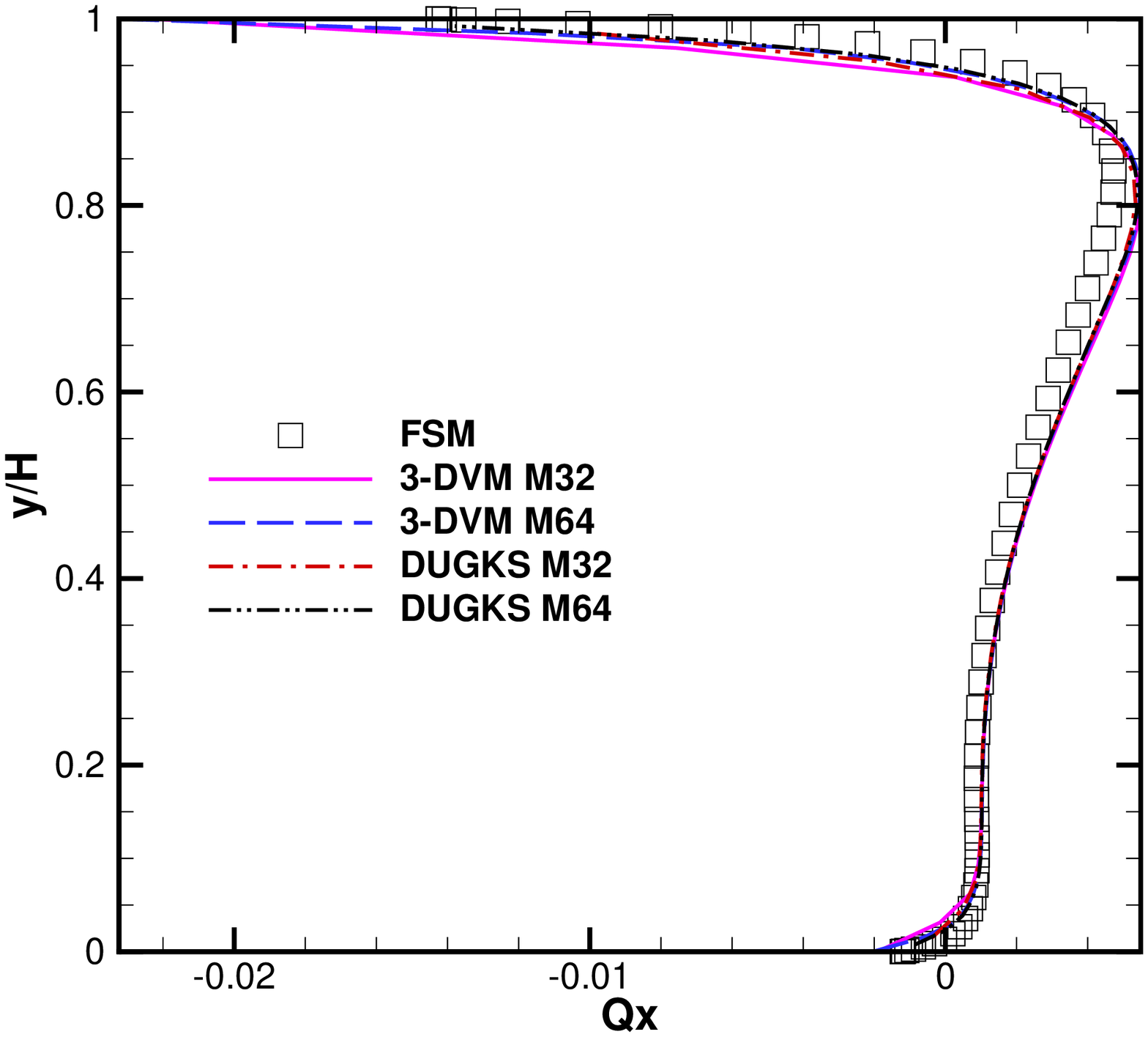}
\caption{$Qx$}
\label{subfig:re10_qx}
\end{subfigure}~
\begin{subfigure}[b]{0.5\textwidth}
\includegraphics[width=\textwidth]{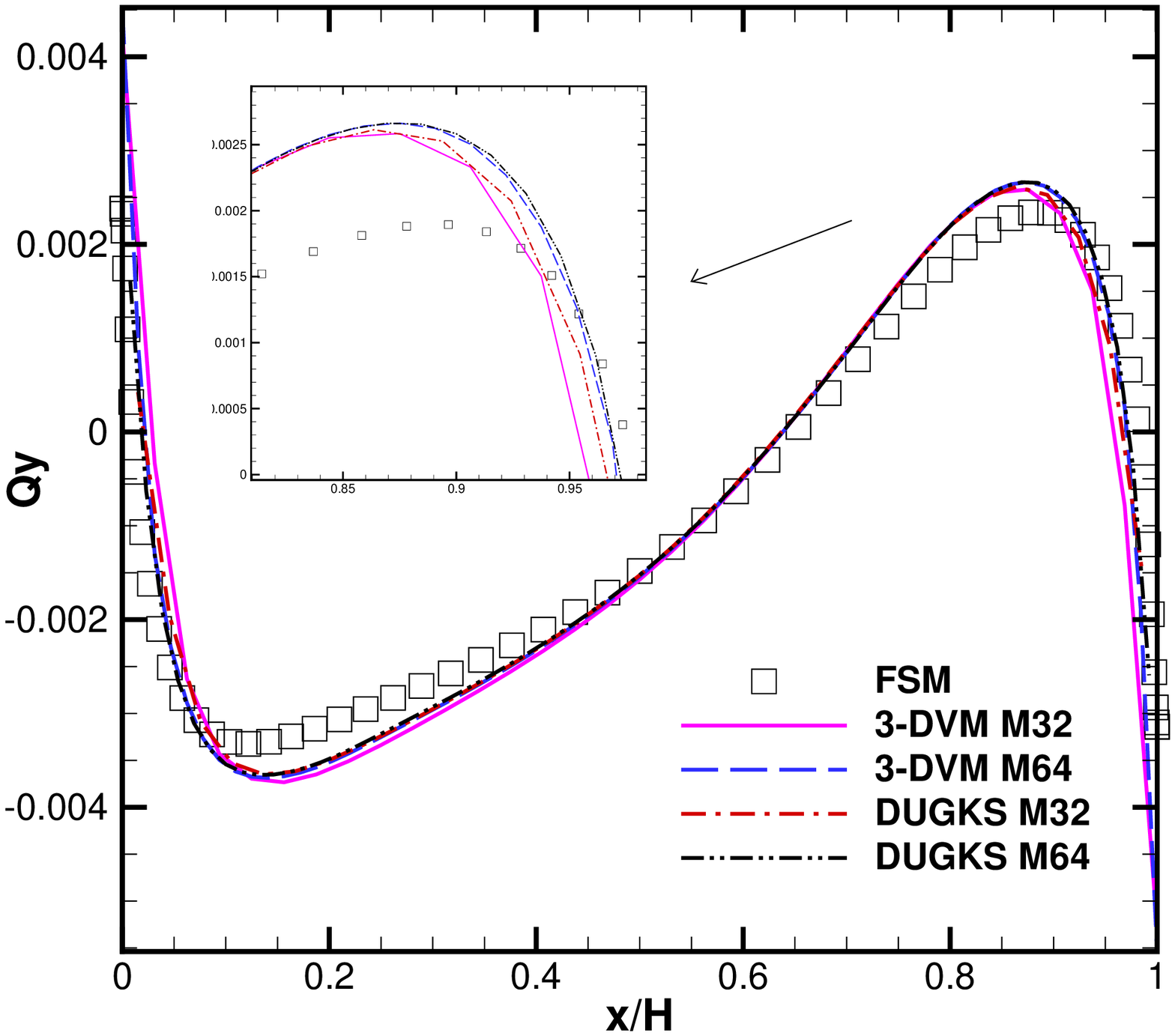}
\caption{$Qy$}
\label{subfig:re10_qy}
\end{subfigure}\\
\caption{
 The results of the cavity flow at $Kn=0.0259$ ($Re=10$): (\textit{a}) U-velocity along the vertical centerline,  (\textit{b}) $V$-velocity along the horizontal centerline,
 (\textit{c}) heat flux $Qx$ along the vertical centerline and (\textit{d}) heat flux $Qy$ along the horizontal centerline.}
\label{fig:cavity_re10}
\end{figure}

\begin{figure}[!htb]
\centering
\begin{subfigure}[b]{0.5\textwidth}
\includegraphics[width=\textwidth]{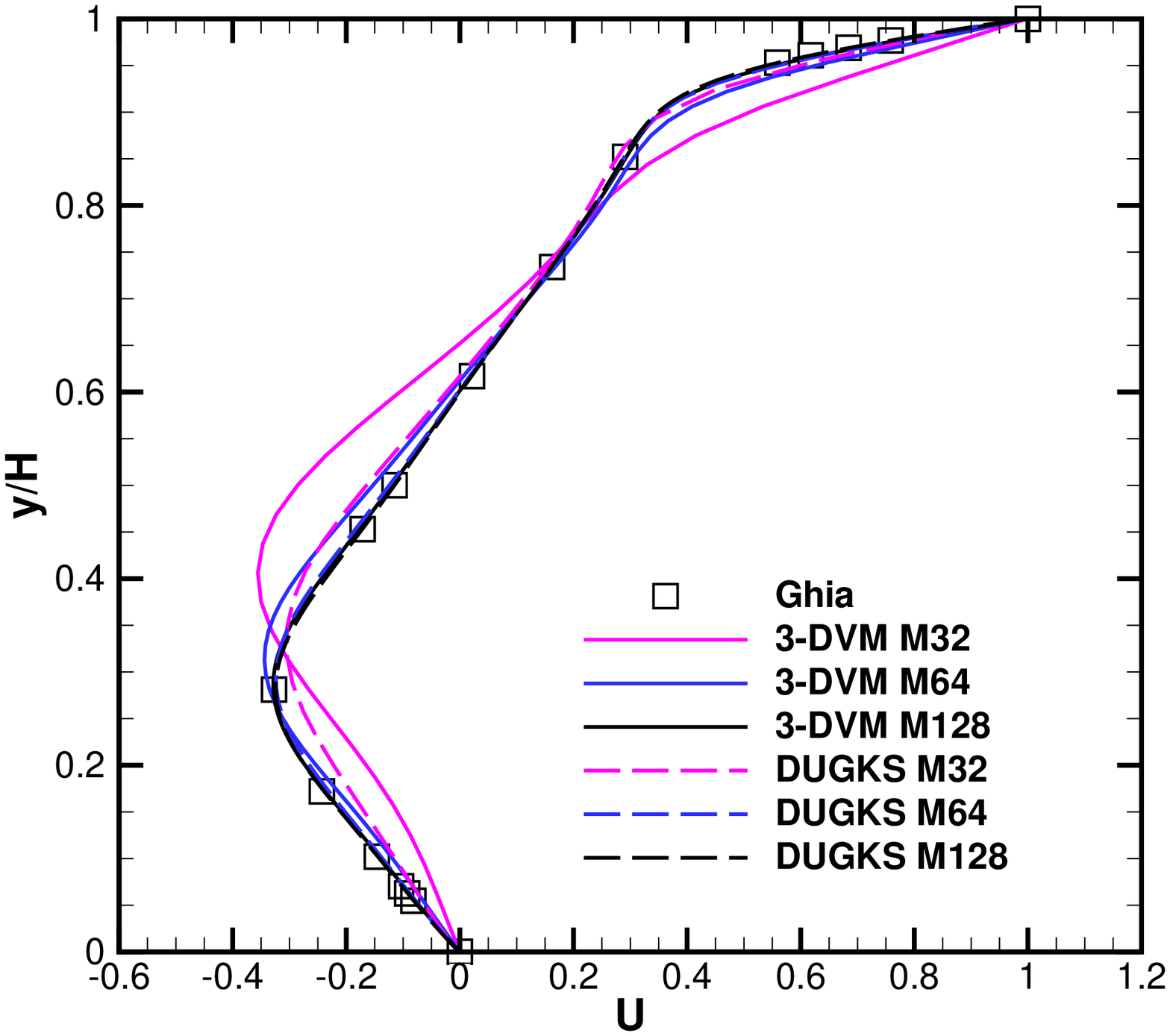}
\caption{$U$}
\label{subfig:re4e2_u}
\end{subfigure}~
\begin{subfigure}[b]{0.5\textwidth}
\includegraphics[width=\textwidth]{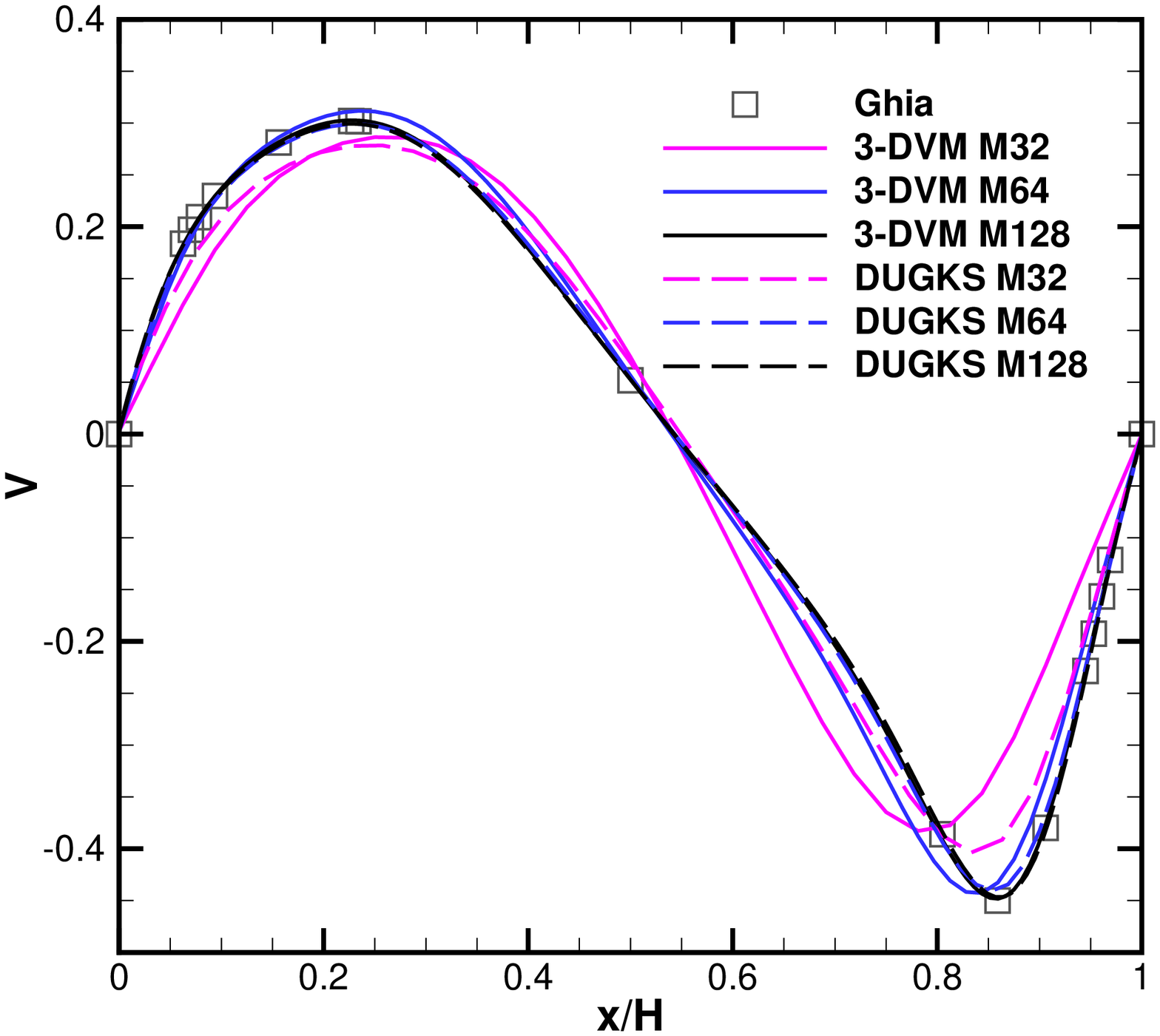}
\caption{$V$}
\label{subfig:re4e2_v}
\end{subfigure}\\
\begin{subfigure}[b]{0.5\textwidth}
\includegraphics[width=\textwidth]{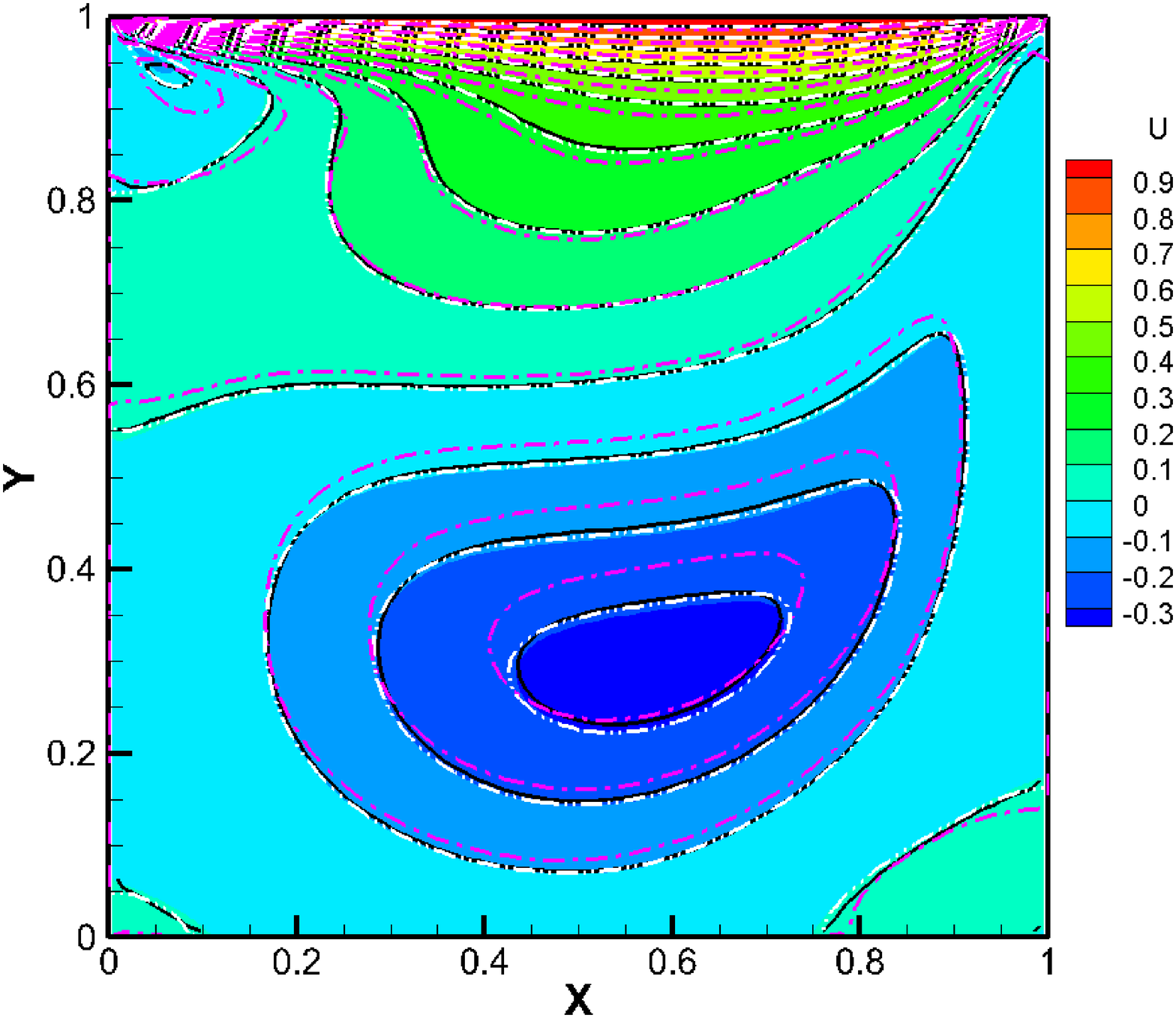}
\caption{$U$-velocity contour}
\label{subfig:re4e2-uc}
\end{subfigure}~
\begin{subfigure}[b]{0.5\textwidth}
\includegraphics[width=\textwidth]{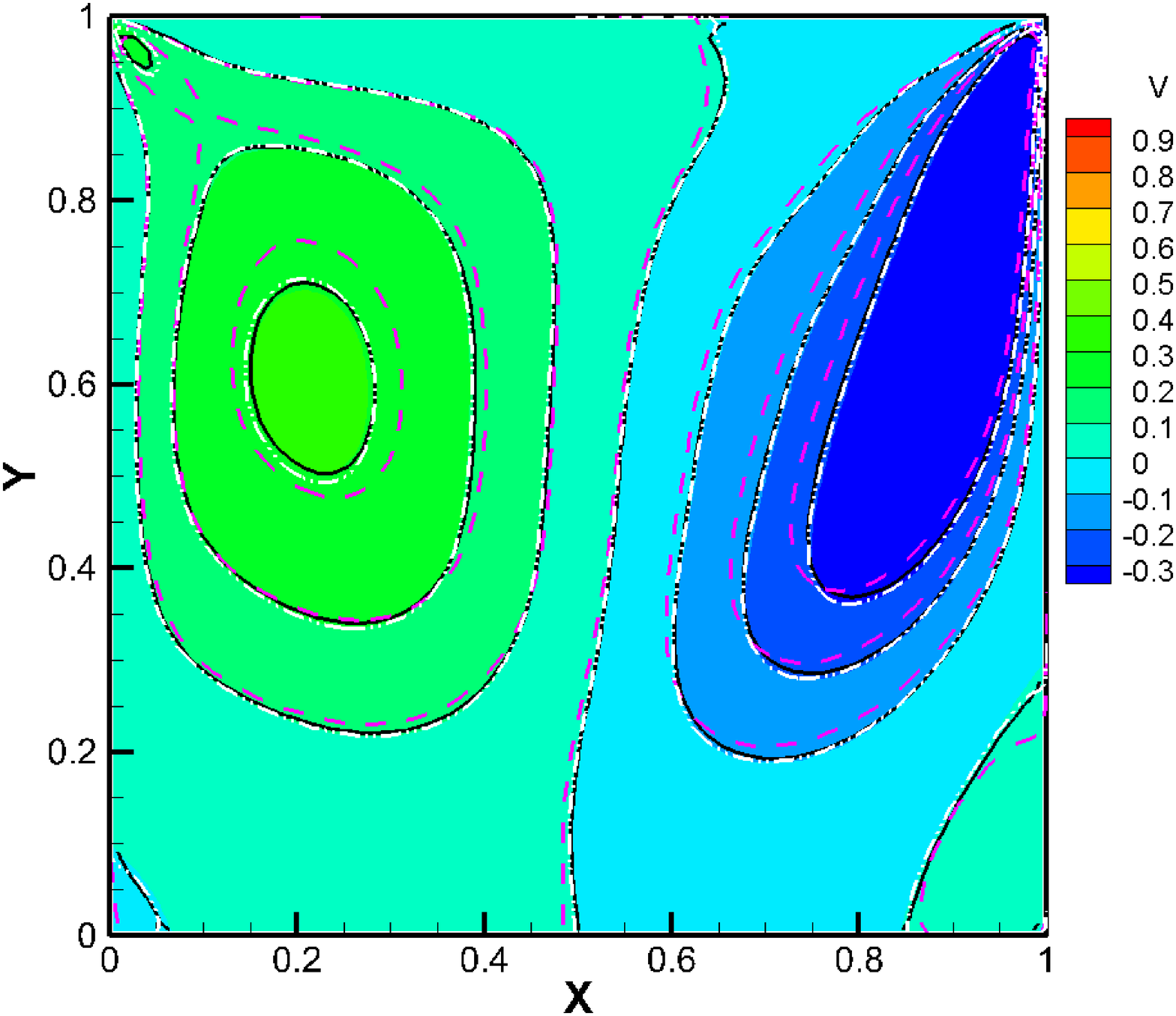}
\caption{$V$-velocity contour}
\label{subfig:re4e2-vc}
\end{subfigure}\\
\caption{
  The results of the cavity flow at $Kn=6.47\times 10^{-4}$ ($Re=400$): (\textit{a}) $U$-velocity along the vertical centerline,
 (\textit{b}) $V$-velocity along the horizontal centerline,
 (\textit{c}) $U$-velocity contour and (\textit{d}) $V$-velocity contour. In the figures (\textit{c}) and (\textit{d}): background, DUGKS with mesh of $128^2$;
  black solid line, DUGKS with mesh $64^2$;
  white dash-double-dotted line, GDVM with mesh of $128^2$;  rose red dash-dotted line, GDVM with mesh $64^2$.}
\label{fig:cavity_re4e2}
\end{figure}

Figure~\ref{fig:cavity_re10} shows the velocity and heat flux along centerlines of the cavity for $Kn=0.0259$ ($Re=10$) in the early slip regime.
It is usually recognized that it is difficult for the traditional DVM in this regime due to requirement of large meshes.
However, we note that the resolved results of GDVM and DUGKS with the fine mesh of $64^2$ are in excellent agreement with each other, although
visible deviations between the GDVM or DUGKS and FSM are still observed for the high-order moment, i.e. heat flux.
This indicates that the mesh requirement of GDVM in such regime is acceptable due to the use of high-order approximation.
In addition, with a coarse mesh of $32^2$, the vertical velocity $V$ and heat flux $Qy$ computed by the DUGKS are slightly better than those from the GDVM.
The same conclusions can be drawn for the transition and free molecular regimes.

The results at $Kn=6.47\times 10^{-4}$ ($Re=400$) in the hydrodynamic regime are also presented, where the benchmark NS
solver solutions are available~\cite{ghia1982high}. Figs.~\ref{subfig:re4e2_u} and \ref{subfig:re4e2_v} show
the horizontal and vertical velocity profiles along the centerlines of the cavity. It is found that with the coarse mesh, the results of DUGKS are
 much better than those of the GDVM. For example, as shown in Fig.~\ref{subfig:re4e2_u}, with the mesh of $32^2$, the GDVM
underestimates the $U$-velocity  boundary layer adjacent to the top wall, whereas the DUGKS can accurately capture this velocity boundary
layer with such coarse resolution. This indicates that the GDVM is more dissipative than the DUGKS. In addition, we also observe that the DUGKS is not so
sensitive to mesh resolutions as the GDVM. This is because that even in this regime DUGKS still preserves the second-order spatial
accuracy~\cite{guo2013discrete,guo2015discrete}. Similar observations can be obtained from Figs.~\ref{subfig:re4e2-uc} and \ref{subfig:re4e2-vc}.
In these two figures, we, respectively, plot the $U$ and $V$ velocity distributions on different mesh resolutions; the well-resolved results of DUGKS
with the finest mesh of $128^2$ are regarded as the reference solutions. It is observed that the results of GDVM with a mesh of $64^2$ clearly
deviate from the reference solutions, particularly around the cavity corners and vortex centers, while the DUGKS with the same mesh can adequately resolve the flow field.
This is consistent with the analysis in Sec.~\ref{analysis} that the DUGKS is more accurate than the GDVM in the continuum regime.

\begin{figure}[htp]
\centering
\begin{subfigure}[b]{0.5\textwidth}
\includegraphics[width=\textwidth]{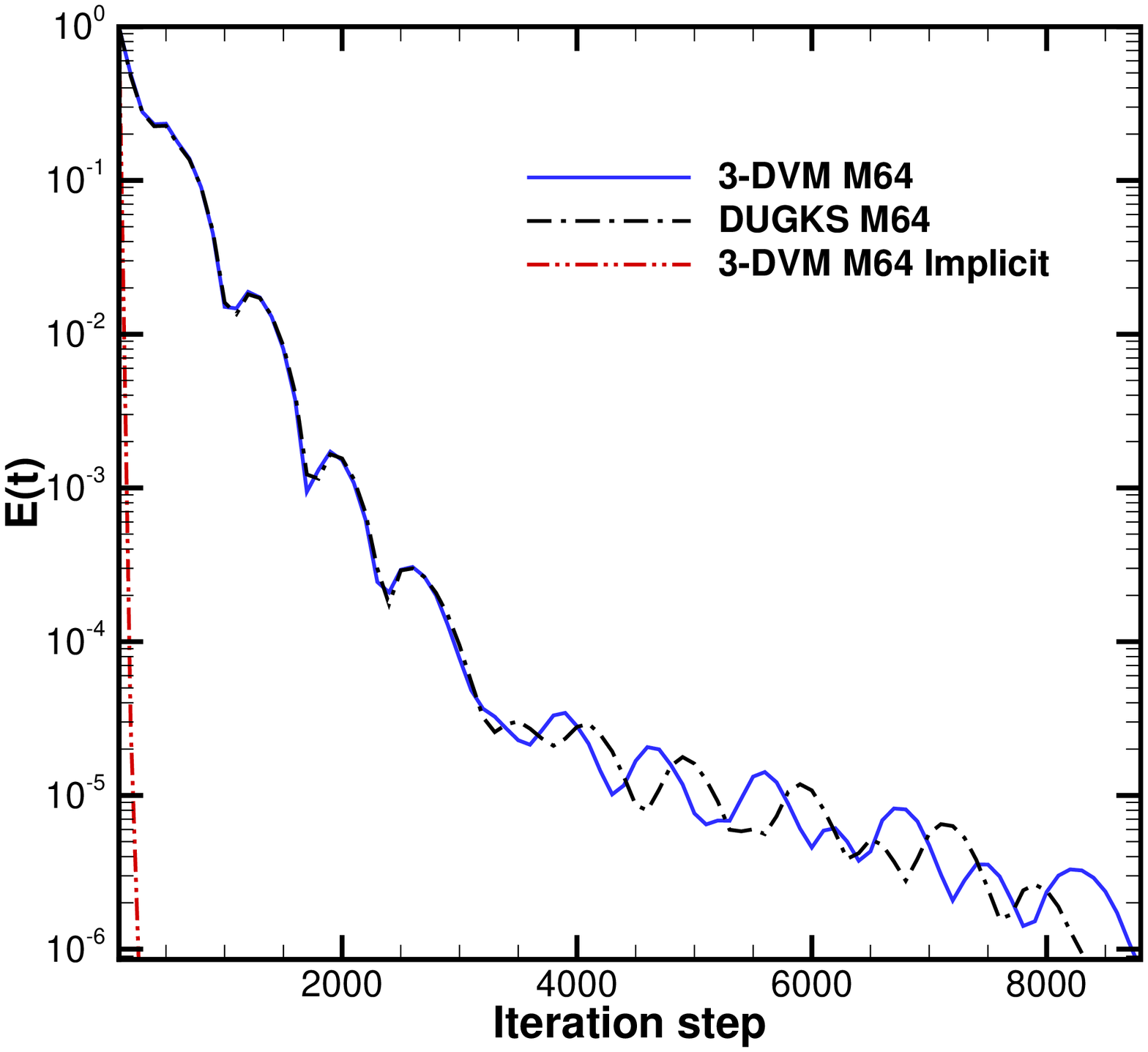}
\caption{$Kn=10$}
\label{subfig:err_kn10}
\end{subfigure}~
\begin{subfigure}[b]{0.5\textwidth}
\includegraphics[width=\textwidth]{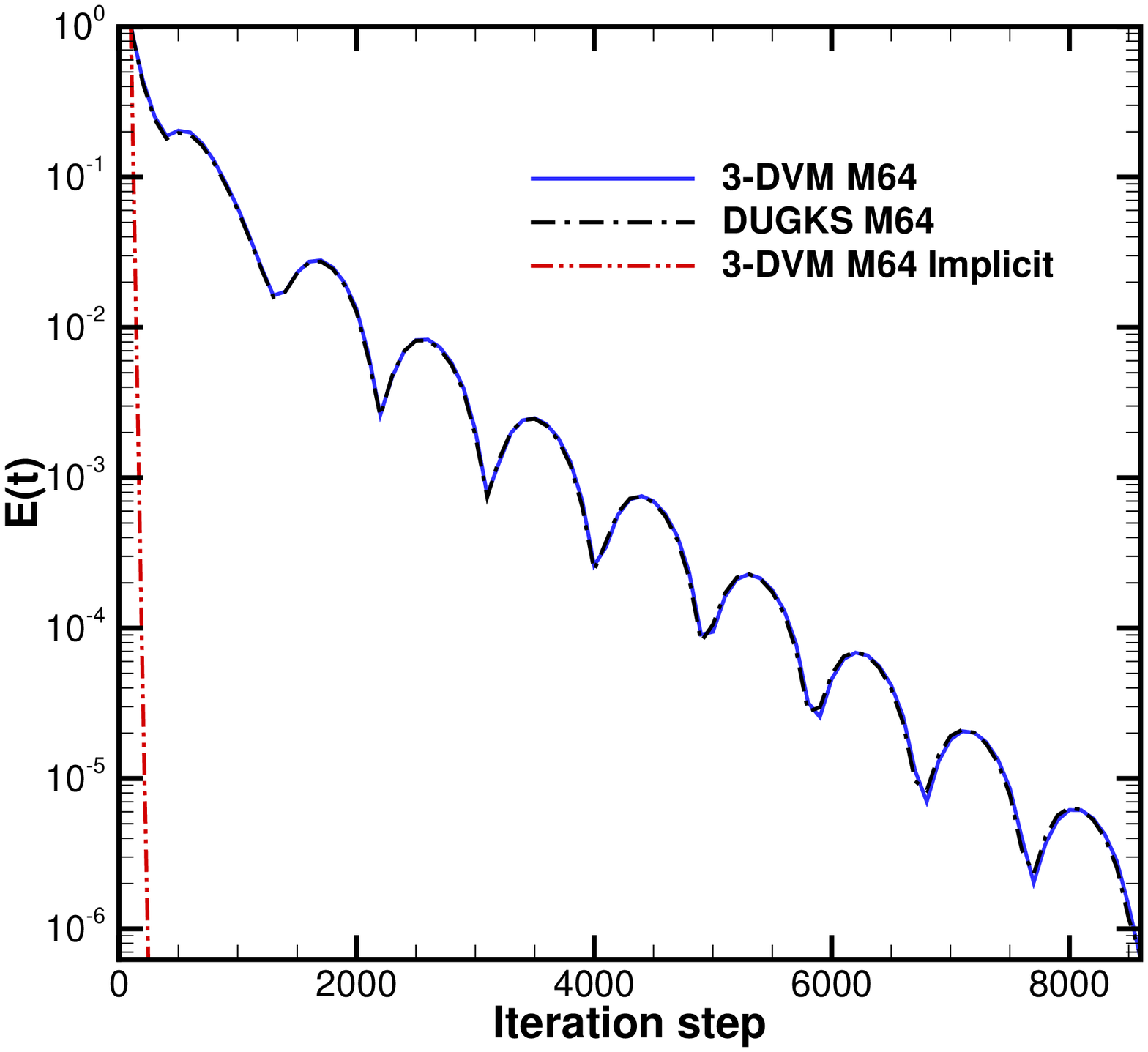}
\caption{$Kn=0.1$}
\label{subfig:err_kn01}
\end{subfigure}\\
\begin{subfigure}[b]{0.5\textwidth}
\includegraphics[width=\textwidth]{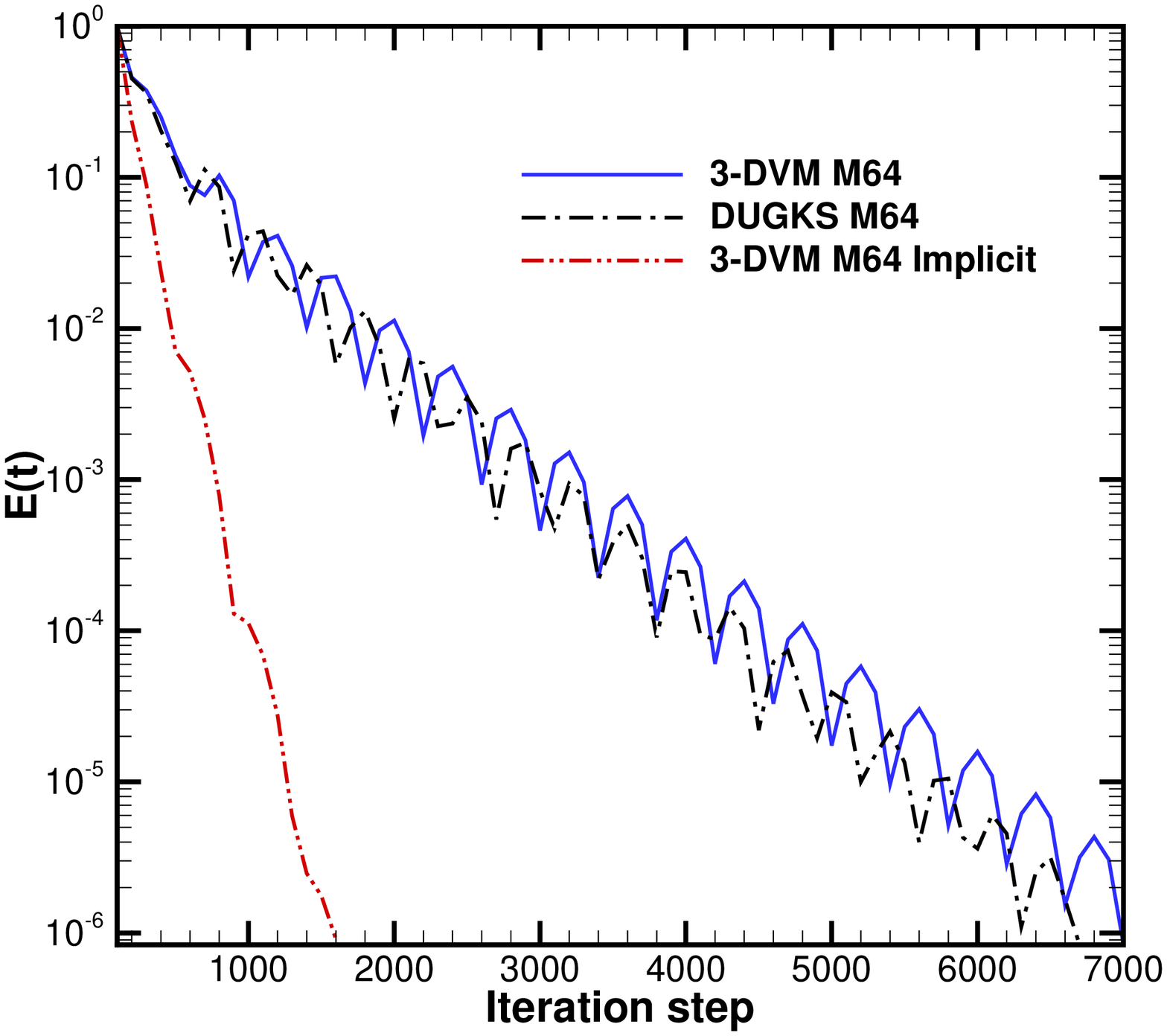}
\caption{${Kn}=2.59\times 10^{-2},$ $Re=10$}
\label{subfig:err_re10}
\end{subfigure}~
\begin{subfigure}[b]{0.5\textwidth}
\includegraphics[width=\textwidth]{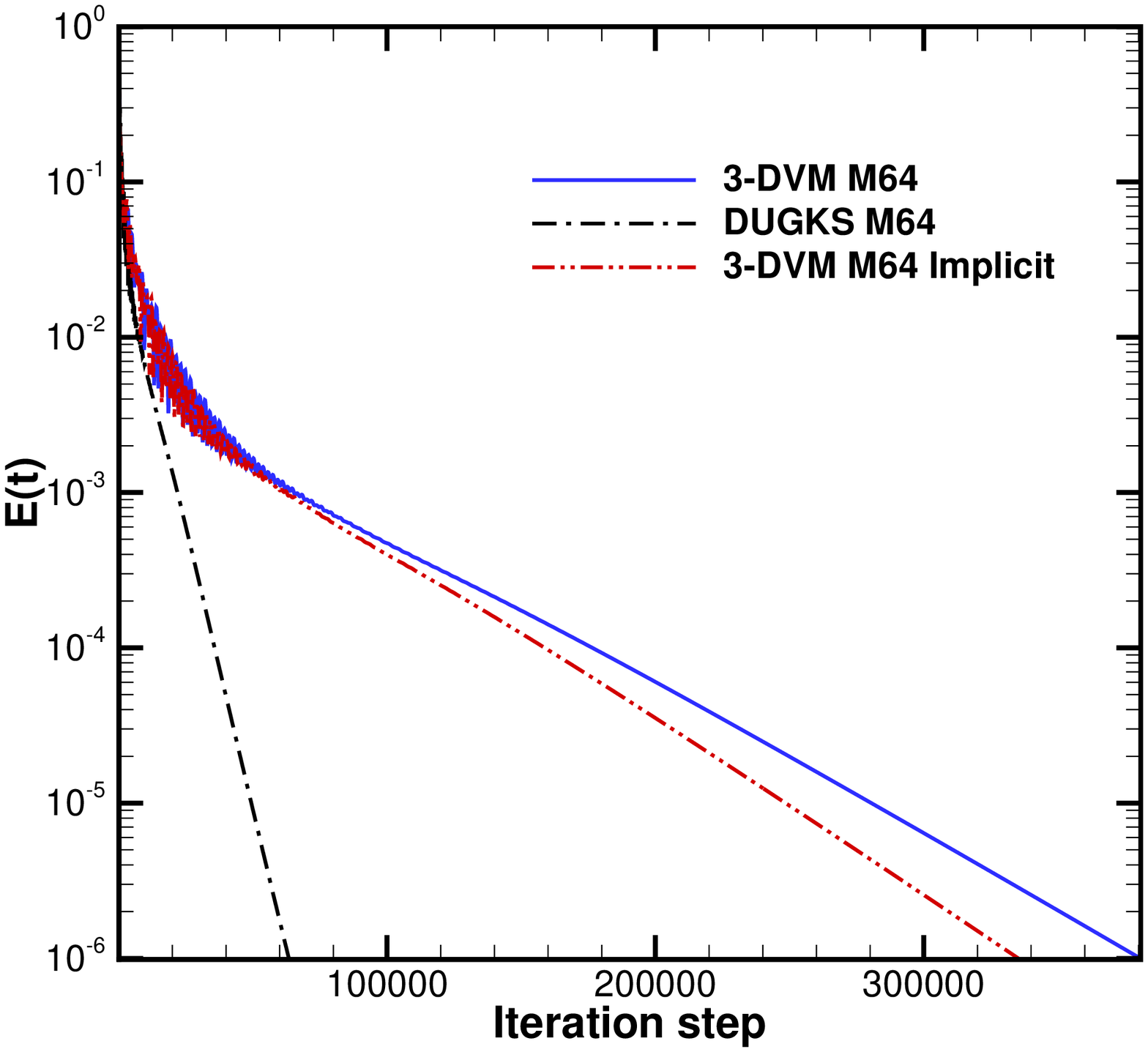}
\caption{${Kn}=6.47\times 10^{-4},$ $Re=400$}
\label{subfig:err_re4e2}
\end{subfigure}\\
\caption{
 Error evolutions defined by Eq.~\eqref{work_state} at different Knudsen numbers: (\textit{a}) $Kn=10$,  (\textit{b}) $Kn=0.1$,
 (\textit{c})  $Kn=2.59\times 10^{-2}$ $(Re=10)$ and (\textit{d}) $Kn=6.47\times 10^{-4}$ $(Re=400)$. Note that for the implicit GDVM, the error is estimated at one time step.}
\label{fig:err_evo}
\end{figure}

Distinct algorithm design of the GDVM and DUGKS may lead to different convergent processes. Fig.~\ref{fig:err_evo} depicts evolutions of the
relative global error defined by Eq.~\eqref{work_state} at different Knudsen numbers. In addition to the results with the
same CFL numbers, the results of GDVM with a large CFL number up to $\eta=10^4$, say implicit GDVM, are also included.
As we can see from Figs.~\ref{subfig:err_kn10} and \ref{subfig:err_kn01}, error evolutions of both methods in the transition and free molecular regimes are
almost identical to each other. However, when approaching to the slip and hydrodynamic regimes,
as shown in Figs.~\ref{subfig:err_re10} and \ref{subfig:err_re4e2}, the convergence rate of DUGKS is apparently faster than that of the GDVM.
Furthermore, we also note that the implicit GDVM converges about two orders of magnitude
faster than the GDVM and DUGKS in highly rarefied regimes, while in the continuum region, as shown in Fig.~\ref{subfig:err_re4e2}, the convergence
rate of DUGKS turns to be two times faster than that of GDVM as well as the implicit GDVM.

\begin{table}[t]
\centering
\caption{\label{tablea} The total CPU time costs (in minute) of the GDVM and DUGKS when the results satisfy the stead-state criterion Eq.~\eqref{work_state}
on the mesh of $64^2$. The results of the implicit GDVM  are also included. Note that the convergency criteria for implicit GDVM is measured at one time step. }
\begin{tabular}{c c c c c c c c c c}
  \hline
   \hline
   $Kn$      & $6.47\times 10^{-4}$ & $0.0259$ & $0.1$ & $1$   & $10$ \\
  \hline
   \hline
  $t_{DUGKS}$                   & 6.06    & 2.35     & 41.52  & 272.93 & 503.31 \\
  $t_{GDVM}$                    & 13.35   & 1.51     & 18.4   & 121.35 & 236.15 \\
$t_{implicit ~GDVM}$             &11.92    & 0.20     & 0.35   & 1.2    & 4.01 \\
$t_{DUGKS}/t_{GDVM}$            & 0.51    & 1.55     & 2.25   & 2.24   & 2.13 \\
$t_{DUGKS}/t_{implicit~ GDVM}$   & 0.46    & 11.75    & 118.62 & 227.43 & 125.51 \\
\hline
\hline
\end{tabular}
\end{table}

\begin{table}[t]
\centering
\caption{\label{tableb} The total CPU time costs (in minute) of the implicit GDVM and DUGKS when the results are well resolved. The convergency criteria for implicit GDVM is measured at one time step. }
\begin{tabular}{c c c c c c c c c c}
  \hline
   \hline
   $Kn$      & & & & & $6.47\times 10^{-4}$ & $0.0259$ & $0.1$ & $1$   & $10$ \\
  \hline
   \hline
  $t_{DUGKS}$                   & & & & & 6.06    & 0.31     & 6.17  & 35.7 & 58.38 \\
$t_{implicit~GDVM}$             & & & & &48.68    & 0.20     & 0.35   & 1.2    & 4.01 \\
$t_{DUGKS}/t_{implicit~GDVM}$   & & & & & 0.12    & 1.55    & 17.63 & 29.75 & 14.56 \\
\hline
\hline& & & & 
\end{tabular}
\end{table}

In addition to accuracy, the computational efficiencies of GDVM and DUGKS are also measured. Firstly, we compare the CPU time cost of
each iteration. For a fair comparison, the time step is set to be identical in the GDVM and DUGKS. For the case of $Kn=0.1$ with $64^2$ mesh points,
the CPU time costs within one time step are 0.1283s and 0.2965s for the GDVM and DUGKS, respectively, which indicates that the GDVM is about one time
faster than the DUGKS for each iteration. According to our analysis in Sec.~\ref{analysis}, this result is not surprising as the DUGKS with a finite-volume
formulation computes more equilibrium state functions than the GDVM with a finite difference one.

However, as shown in Fig.~\ref{fig:err_evo}, there are different convergence rates for the GDVM and DUGKS in various regimes,
which will lead to different time costs to achieve a convergent solution. This assessment includes not only the time cost of
these two methods with a same time step, but also the implicit GDVM. Table~\ref{tablea} presents the total CPU time costs to attain the
steady-state (Eq.~\eqref{work_state}) in various regimes. Note that for the implicit GDVM, the error estimation is performed for each iteration. As expected, in the transition and free molecular regimes,
the GDVM is about one time faster than the DUGKS, whereas as $Kn$ decreases, the GDVM becomes slower than the DUGKS due to the faster
convergence rate of DUGKS. Moreover, it is interesting to note that although the efficiency of
implicit GDVM is improved by two orders of magnitude in highly rarefied regimes, it is still about one time slower than the DUGKS
in the hydrodynamic regime.

It should be noted that the above efficiency comparisons are made based on the same mesh for two methods. As shown, the GDVM requires $64^2$ mesh points to obtain the resolved results for the flows from early slip to highly rarefied regimes, while for the DUGKS, it only needs a coarser mesh of $32^2$. Likewise, for the continuum flow, the mesh requirements for the GDVM and DUGKS are $128^2$ and $64^2$, respectively. Therefore, the DUGKS
can achieve accurate results with coarser meshes in comparison with the GDVM.  Consequently, as shown in table~\ref{tableb}, to achieve the well-resolved results, the DUGKS is about one order of magnitude faster than the implicit GDVM in the
continuum region, and vice versa in the highly rarefied regimes.

In addition, concerning the parallel computational aspect, it is straightforward for both GDVM and DUGKS to decompose
molecular velocity space. However, if the physical domain decomposition is of interest, the parallel implementation of this implicit GDVM is considerably
more challenging than the DUGKS counterpart.


\section{Conclusions}
\label{conclusion}

The main objective of this work is to quantify the computational performance of different DVMs,
so that researchers may choose the most appropriate method for their applications.
Our results show that both the GDVM and DUGKS can accurately reproduce the results in all the flow regimes, provided that the
mesh resolution is sufficient. Meanwhile, it is found that the DUGKS is less dissipative and consequently requires much smaller number of meshes than the GDVM,
especially in the continuum and near continuum regimes. In the GDVM, the convection term of the kinetic model is approximated by the upwind scheme with
the underlying assumption of molecular free streaming
between two grid points, while in the DUGKS the collision and transport processes are coupled physically by using the discrete characteristic solution of
the kinetic equation. Therefore, even with a third-order discretization, the GDVM is not as accurate as the second-order DUGKS,
particularly in such regimes where intensive molecular collisions take place.

The efficiency and convergence rate of the GDVM and DUGKS are also compared. Our results showed that with the same mesh for each iteration, the CPU time cost of
the DUGKS is about twice that of the GDVM, which is not surprising that the finite-volume DUGKS computes more equilibrium
state distribution functions when compared to the GDVM with a finite-difference formulation. In addition, when using the same time step and spatial mesh,
the GDVM and DUGKS showed similar convergence rates from the free molecular to early slip regimes, so that the GDVM is about twice as fast as
the DUGKS; when using a large time step, the implicit GDVM is faster than the explicit DUGKS by about two orders of magnitude.
However, as the flow approaches to the hydrodynamic regime in which molecular collisions dominate, the DUGKS converges faster,
consequently, it turns out to be twice as fast as the implicit GDVM. It should be noted that in order to achieve results in reasonable accuracy,
the DUGKS requires fewer mesh points than the GDVM, therefore, the overall computational efficiency of DUGKS can be improved by one order of magnitude.

In summary,
the DUGKS is preferable for flow problems involving different flow regimes, while if only the steady-state solution of highly rarefied
flows is of interest, the implicit GDVM, which can boost the GDVM convergence rate by two orders of magnitude, is a better choice.

\section*{ACKNOWLEDGMENT}
This work is financially supported by the UKs Engineering and Physical Sciences Research Council (EPSRC) under grants EP/M021475/1, EP/L00030X/1 and National Natural Science Foundation 
of China (Grant No. 51125024).

\section*{References}

\bibliographystyle{elsarticle-num}
\bibliography{DVM-DUGKS}
\end{document}